\documentclass[fleqn,usenatbib]{mnras}

\usepackage{newtxtext,newtxmath}
\usepackage{lipsum}
\usepackage[T1]{fontenc}
\usepackage{longtable}
\DeclareRobustCommand{\VAN}[3]{#2}
\let\VANthebibliography\thebibliography
\def\thebibliography{\DeclareRobustCommand{\VAN}[3]{##3}\VANthebibliography}

\usepackage{graphicx}	
\usepackage{amsmath}	
\usepackage{pdflscape}

\usepackage{supertabular}
\usepackage{microtype} 
\usepackage{enumitem}

\usepackage{etoolbox, siunitx}
\robustify\bfseries

\usepackage{booktabs} 
\usepackage[normalem]{ulem} 

\DeclareSIUnit\au{AU}
\DeclareSIUnit\Rsun{R_\odot}
\DeclareSIUnit\Rjup{R_\text{Jup}}
\DeclareSIUnit\Msun{M_\odot}
\DeclareSIUnit\Mjup{M_\text{Jup}}
\DeclareSIUnit\gyr{Gyr}
\DeclareSIUnit\ppt{ppt}
\DeclareSIUnit\ppm{ppm}

\usepackage{xcolor}

\title[EBLM XVII. Tidal Synchronization \& Circularization]{EBLM XVII - Tidal Synchronization and Circularization in Tight Stellar Binaries}

\author[Sethi et al.]{%
        Ritika Sethi$^{1,2*}$$^{\href{https://orcid.org/0000-0002-6576-3346}{\includegraphics[scale=0.5]{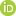}}}$,
        David V. Martin$^{3}$$^{\href{https://orcid.org/0000-0002-7595-6360}{\includegraphics[scale=0.5]{orcid.jpg}}}$,
        Adrian Barker$^{4}$$^{\href{https://orcid.org/0000-0003-4397-7332}{\includegraphics[scale=0.5]{orcid.jpg}}}$,
        Pierre  F. L. Maxted$^{5}$$^{\href{https://orcid.org/0000-0003-3794-1317}{\includegraphics[scale=0.5]{orcid.jpg}}}$,\newauthor
        Amaury H.M.J. Triaud$^{6}$$^{\href{https://orcid.org/0000-0002-5510-8751}{\includegraphics[scale=0.5]{orcid.jpg}}}$,
        Vedad Kunovac$^{7,8}$$^{\href{https://orcid.org/0000-0001-9419-3736}{\includegraphics[scale=0.5]{orcid.jpg}}}$,
        Wata Tubthong$^{3}$$^{\href{https://orcid.org/0000-0002-7907-2634}{\includegraphics[scale=0.5]{orcid.jpg}}}$,
        Alison Duck$^{9,10}$$^{\href{https://orcid.org/0000-0002-4531-6899}{\includegraphics[scale=0.5]{orcid.jpg}}}$,\newauthor
        François Bouchy$^{11}$$^{\href{https://orcid.org/0000-0002-7613-393X}{\includegraphics[scale=0.5]{orcid.jpg}}}$,  
        Stéphane Udry$^{11}$$^{\href{https://orcid.org/0000-0001-7576-6236}{\includegraphics[scale=0.5]{orcid.jpg}}}$\\
\\
$^{1}$Department of Physics, Massachusetts Institute of Technology, Cambridge, MA 02139, USA \\
$^{2}$MIT Kavli Institute for Astrophysics and Space Research, Massachusetts Institute of Technology, Cambridge, MA 02139, USA\\
$^{3}$Department of Physics \& Astronomy, Tufts University, Medford, MA 02155, USA \\
$^{4}$School of Mathematics, University of Leeds, Leeds LS2 9JT, UK \\
$^{5}$Astrophysics Group, Keele University, Keele, Staffordshire, ST5 5BG, UK\\
$^{6}$School of Physics and Astronomy, University of Birmingham, Edgbaston, Birmingham B15 2TT, UK\\
$^{7}$Centre for Exoplanets and Habitability, University of Warwick, Coventry, CV4 7AL, UK\\
$^{8}$Department of Physics, University of Warwick, Coventry, CV4 7AL, UK\\
$^{9}$Jet Propulsion Laboratory, California Institute of Technology, 4800 Oak Grove Drive, Pasadena, CA 91109, USA\\
$^{10}$Department of Astronomy, The Ohio State University, Columbus, OH 43210, USA\\
$^{11}$Observatoire Astronomique de l’Université de Genève, Chemin Pegasi 51, CH-1290 Versoix, Switzerland\\
*rsethi@mit.edu\\
}

\date{First submitted to MNRAS Nov 6 666}

\pubyear{2025}

\begin{document}

\label{firstpage}
\maketitle

\begin{abstract}
Tidal interactions in close stellar binaries are central to their orbital and rotational evolution, making observational tests of theoretical predictions essential for our understanding of the evolution of these, as well as close exoplanetary systems. Such tests require precise measurements of the orbital eccentricity and stellar rotation. The EBLM (Eclipsing Binary Low Mass) survey delivers a homogeneous sample of eclipsing binaries, composed of F/G/K primaries and M-dwarf (or low-mass K-dwarf) secondaries. We analyze 68 unequal mass binaries ($0.1 \leq q \leq 0.6$, where $q$ is the mass ratio), with measurable primary star rotation rates from TESS, and over a decade of radial velocity observations. This sample probes the critical regime where tidal effects are expected to transition between being efficient and inefficient. We find that $\sim$75\% of our sample has circularized, with eccentric systems confined to $P_{\rm orb} \gtrsim 3$ days, with modest eccentricities ($e < 0.25$). Roughly $\sim$78\% of our sample is synchronized, with nearly all binaries within a 3-day orbital period residing in a well-defined “synchronization zone”. Beyond this, a minority of asynchronous systems persist, which cannot be easily explained by our application of current tidal mechanisms or by differential rotation.
\end{abstract}


\section{Introduction}\label{sec:introduction}
Stellar binaries are ubiquitous in the universe, with roughly half of the solar-type stars existing in binary or multi-star systems \citep{1991A&A...248..485D, 2013ARA&A..51..269D, 2023ASPC..534..275O}. Close stellar binaries, in particular, serve as laboratories for studying tidal physics. 

The differential gravitational force exerted by one star on its companion in a binary system gives rise to a variety of tidal effects, including stellar shape distortions and the generation of tidal torques as a result of the dissipation of tidal flows \citep[e.g.][]{1980A&A....Hut,2005ASPC..333...95M,2006ApJ...653..621M}. When the tidal potential is time-varying, angular momentum is exchanged between stellar spins and the orbit when tidal flows are dissipated. This process drives the system toward a state of tidal equilibrium  \citep[e.g.][]{1980A&A....Hut,BO2009,2010A&A...516A..64L}. Specifically, tidal interactions tend to evolve binary systems toward three principal dynamical end states:
\begin{enumerate}
 \item {\bf Tidal circularization}, where tidal dissipation reduces orbital eccentricity, eventually leading to a circular orbit.
\item {\bf Synchronization (i.e. tidal locking)}, where both stellar rotation periods ($P_{\rm rot}$) match the orbital period ($P_{\rm orb}$).
\item {\bf Spin-orbit alignment}, where the orbital plane aligns with the rotation axis of both stars
\end{enumerate}
This final equilibrium state of circularity, spin-orbit synchronization and coplanarity is possible if there is sufficient angular momentum and if angular momentum is approximately conserved in a given two-body system \citep[e.g.][]{1980A&A....Hut,2014ARA&A..52..171O,Barker2025}, otherwise the two bodies will merge.

Theoretical models predict that tidal evolutionary timescales depend on a range of stellar and orbital parameters, including the mass ratio ($q = M_{\rm B}/M_{\rm A}$, where the subscript A denotes the more massive star and B denotes the less massive star), primary mass ($M_{\rm A}$), stellar radius, orbital period ($P_{\rm orb}$), and eccentricity ($e$) \citep[e.g.][]{1977A&A....Zahn, 2007A&A...461.1057T,2014ARA&A..52..171O,2016AJ....151..139M, 2018ApJ...866...67T, Baker}. Observationally testing these models presents various challenges. In the context of tidal synchronization, precise measurements of stellar rotation periods are required. Traditionally, stellar rotation has been inferred spectroscopically by measuring line broadening to estimate the projected rotational velocity ($v\sin{i}$). Converting $v\sin{i}$  to the rotation period requires independent estimates of both the stellar radius and the inclination angle ($i_\star$, the angle between the stellar rotation axis and the line of sight), introducing uncertainties and potential biases in the rotation period estimation. The advent of high-precision photometry from TESS and Kepler has transformed this landscape by enabling direct measurement of the true (not projected) rotation period using starspot modulation \citep{2012MNRAS.419.3147A, 2013MNRAS.432.1203M}. However, a caveat to the use of starspots is the role of differential rotation, which can confuse interpretation of the photometric spot modulation. It should also be remembered that both of these methods measure the surface rotation, and not the mean rotation, of a star. 

Additionally, earlier tidal synchronization studies focused mainly on early-type stars with radiative envelopes \citep{2004ApJ...616..562A,2010yCat..74010257K,2023MNRAS.525.5880H,2024A&A...684A..35B}, whereas late-type stars, with convective envelopes, may be dominated by different tidal dissipation mechanisms \citep[e.g.][]{2014ARA&A..52..171O,Barker2025}. \citet{2017AJ....Lurie} analyzed tidal synchronization in late-type Kepler binaries and identified a transition from synchronous to asynchronously rotating systems at an orbital period around 10 days. A more recent study by \citet{2025arXiv250104082H} presented the largest catalog of eclipsing binaries with measured orbital and rotational periods from TESS, broadly confirming the trends observed by \citet{2017AJ....Lurie} and \citet{2021ApJ...912..123J}. However, \citet[][]{2017AJ....Lurie} estimated the primary mass and mass ratio based on photometric colors and eclipse depth ratios, so these values were not very precise. Their sample was also dominated by systems with either extremely small mass ratios (primary eclipse depth <0.1 and no detected secondary eclipse) or nearly equal mass ratios (eclipse depth ratio > 0.7), leaving the small to intermediate mass ratio regime largely unexplored.

With respect to tidal circularization, numerous theoretical models have been developed, yet their predictions are often in conflict with each other \citep{1989A&A...220..112Z,1989ApJ...342.1079G,1989A&A...223..112Z,GD1998,T1998,OL2007,2009ApJ...704..930P,Baker,2022ApJ...927L..36B}, highlighting the need for stronger observational constraints. Previous empirical studies \citep{2005ASPC..333...95M,2011ApJ...733..122D,2022MNRAS.516.6145P,2023MNRAS.518.2885C} have provided valuable insights but were often limited by small or heterogeneous samples.

As a result, a detailed analysis of tidal synchronization and circularization in low-mass binaries was previously unfeasible, primarily due to the lack of a well-characterized sample spanning small to intermediate mass ratios. The EBLM (Eclipsing Binary Low Mass) project, initiated in 2010, has since confirmed hundreds of binaries consisting of an F, G, or K-type primary star and an M-dwarf (or low-mass K-dwarf) secondary. These binaries were originally identified as false-positive hot Jupiter candidates in the WASP (Wide Angle Search for Planets) survey. Through more than a decade of spectroscopic follow-up, the EBLM program has provided a homogeneous, well-characterized sample of binaries with small to intermediate mass ratios ($q < 0.6$) \citep{2013A&A...549A..18T, 2014A&A...572A..50G, 2023MNRAS.521.6305D, 2023MNRAS.519.3546S, 2023Univ....9..498M, 2024MNRAS.535.3343M, 2025MNRAS.tmp..257F, 2025MNRAS.540.2914S}. The largest published release \citep{2017A&A...EBLM4} includes 118 systems with precise eccentricity measurements. 

In this paper, we analyze 68 systems from the EBLM sample whose primary stars have measured rotation periods from \citet{2024MNRAS.529.4442S} and most eccentricity measurements are taken from \citet{2017A&A...EBLM4}. Our focus is on the intermediate to small mass ratio regime ($0.1 \leq q \leq 0.6$) that was not explored by \citet{2017AJ....Lurie} for late-type stars. We used dynamical masses derived from radial velocities \citep[hereafter RVs,][]{2017A&A...EBLM4}, which offer far greater precision than photometric estimates, together with rotation periods measured from TESS starspot modulations \citep{2024MNRAS.529.4442S}. This dataset is ideally suited for probing the transition between synchronized and asynchronous regimes, especially for either fully convective stars or those with convective envelopes. It allows us to identify the threshold where tidal effects are inferred to become significant in such systems. Our work provides one of the most comprehensive empirical tests of tides in low-mass binaries to date, yielding critical constraints for testing tidal theories and advancing our understanding of the formation and early evolution of close binary systems.
A key caveat, however, is that our photometric rotation measurements pertain only to the primary stars. The rotation states of the secondaries cannot be determined from the available photometric data, and while the secondary star may also be synchronized or spin–orbit aligned, we cannot assess this with current observations.

The paper is structured as follows: in \S~\ref{sec:sample} we describe the sample selected for this study. \S~\ref{sec:methods} outlines our methodology, including the calculation of circularization and synchronization timescales for the primary stars using current tidal models. \S~\ref{sec:results} presents the main results and our interpretations and provides short notes on selected outliers and systems with interesting properties that represent key targets for future studies. Finally, we conclude in \S~\ref{sec:conclusion}.

\begin{figure}
    \centering
    \includegraphics[width=0.48\textwidth]{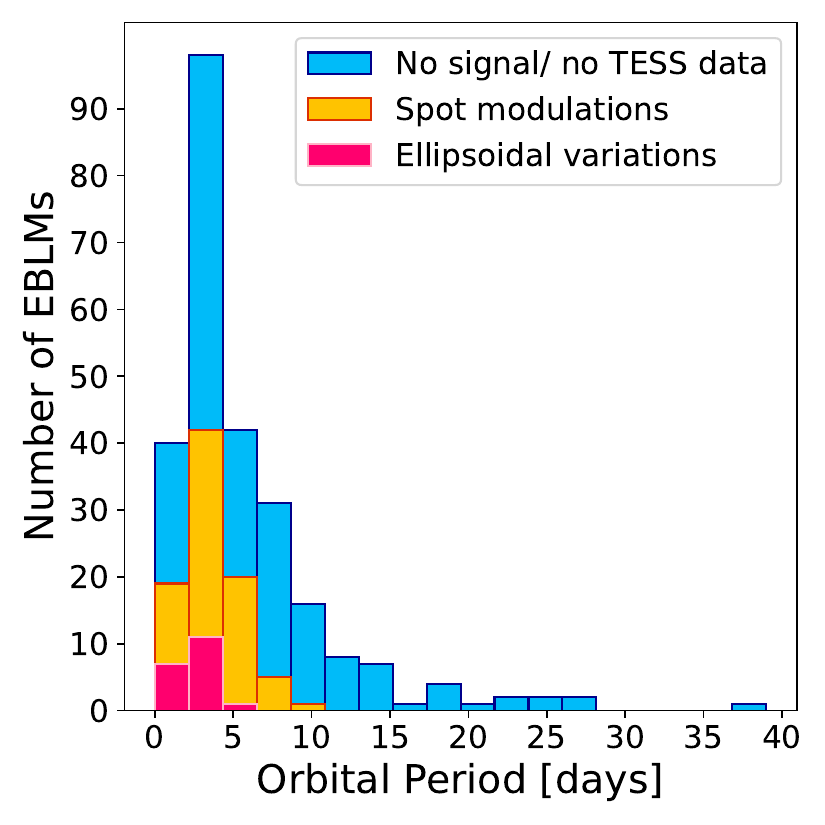}
    \caption{Histogram showing the orbital period distribution of binaries from the EBLM catalog. The three histograms are stacked vertically for clarity. The top histogram represents EBLMs for which no rotational signature was detected, the middle one highlights the systems for which \citet{2024MNRAS.529.4442S} measured rotation periods from TESS light curves exhibiting out-of-eclipse variability due to starspots, and the bottom histogram shows the ellipsoidal variables discovered by \citet{2024MNRAS.529.4442S}.}
    \label{fig:histogram}
\end{figure}

\section{Sample Selection} \label{sec:sample}
Our goal is to study tidal synchronization and circularization within a sample of eclipsing binaries with $q < 0.6$. To study synchronization, we compare the orbital period, $P_{\rm orb}$, and the stellar rotation period, $P_{\rm rot}$. The binaries analyzed here are drawn from the EBLM sample with measured rotation periods from \citet{2024MNRAS.529.4442S}. For the majority of our sample, we use the orbital period published in \citet{2017A&A...EBLM4}. For the few EBLM targets that were not in \citet{2017A&A...EBLM4}, we use their respective discovery paper within the EBLM series. The rotation periods of the primary star come from periodic modulations in the TESS photometry. Since starspots and faculae have different temperatures and brightnesses than the surrounding stellar surface, they produce periodic flux variations in a star’s light curve as the star rotates. The primary star in these systems is at least 40 times brighter than its companion, ensuring that the observed starspot modulation arises from the rotation of the primary star. For the secondary to contribute a detectable photometric signal, it would need to be extremely active, producing substantial spot coverage even in its brightest state. While such contamination cannot be completely ruled out, it is expected to affect only a very small fraction of systems. Moreover, we find no evidence of secondary-induced variability in the observed light curves. This brightness contrast provides confidence that the measured spot modulation likely comes the primary’s rotation. However, a limitation of this approach is that the rotation period of the secondary star cannot be determined, restricting any tidal synchronization analysis to the primary.

A further consideration is that periodic out-of-eclipse variability can also arise from ellipsoidal variations, produced by tidal distortions that deform the star from a spherical to an ellipsoidal shape. However, ellipsoidal variability is readily distinguished from rotational modulation, as it produces a coherent signal with minima coinciding with the primary and secondary eclipses. \citet{2024MNRAS.529.4442S} implemented a multi-step procedure to differentiate between these sources of variability. Their analysis included preprocessing the TESS light curves, identifying periodicities using both autocorrelation functions (ACF) and Lomb–Scargle periodograms, and performing a two-sinusoid fit to classify the source of each signal as spot or ellipsoidal variation. Systems exhibiting spot-induced modulation were retained for rotation period analysis, while those showing ellipsoidal variability were classified separately. Among the 208 EBLM targets cataloged up to June 2022, \citet{2024MNRAS.529.4442S} successfully detected spot modulations in the TESS light curves of 69 systems, and measured their rotation periods. Additionally they discovered 17 ellipsoidal variables. This distribution is shown in the histogram in Figure \ref{fig:histogram}. 

Our sample consists of 68 systems from \citet{2024MNRAS.529.4442S} and is restricted to EBLMs with orbital periods $P_{\rm orb} < 10$ days--the period around which previous studies \citep[e.g.][]{2017AJ....Lurie} observed a transition from synchronous to asynchronous rotation. This is partly because longer stellar rotation periods are difficult to detect from TESS photometry, as the spacecraft’s 13.7-day orbit introduces systematics that can be mistaken for true astrophysical variability. Nevertheless, since our binaries are of small mass ratio, our period range covers the transition regime between effective and ineffective tides according to \citet{2017AJ....Lurie}. However, extending rotation period measurements to longer-period systems in future work is encouraged as it may further refine constraints on tidal synchronization at wider separations.

To study circularization, we use precisely measured orbital eccentricities largely taken from \citet{2017A&A...EBLM4}, who measured a median eccentricity precision of 0.0025. To connect our observations of tidal synchronization and circularization to theoretical expectations, we also require the masses of both stars and the radius of the primary star. These also were largely taken from \citet{2017A&A...EBLM4}. A complete list of measured rotation periods and corresponding orbital parameters for the EBLM systems analyzed in this paper is provided in Table~\ref{Tab:Rot per}.

\section{Methods of Analysis} \label{sec:methods}

\begin{figure*}
    \centering
    \includegraphics[width=1\textwidth]{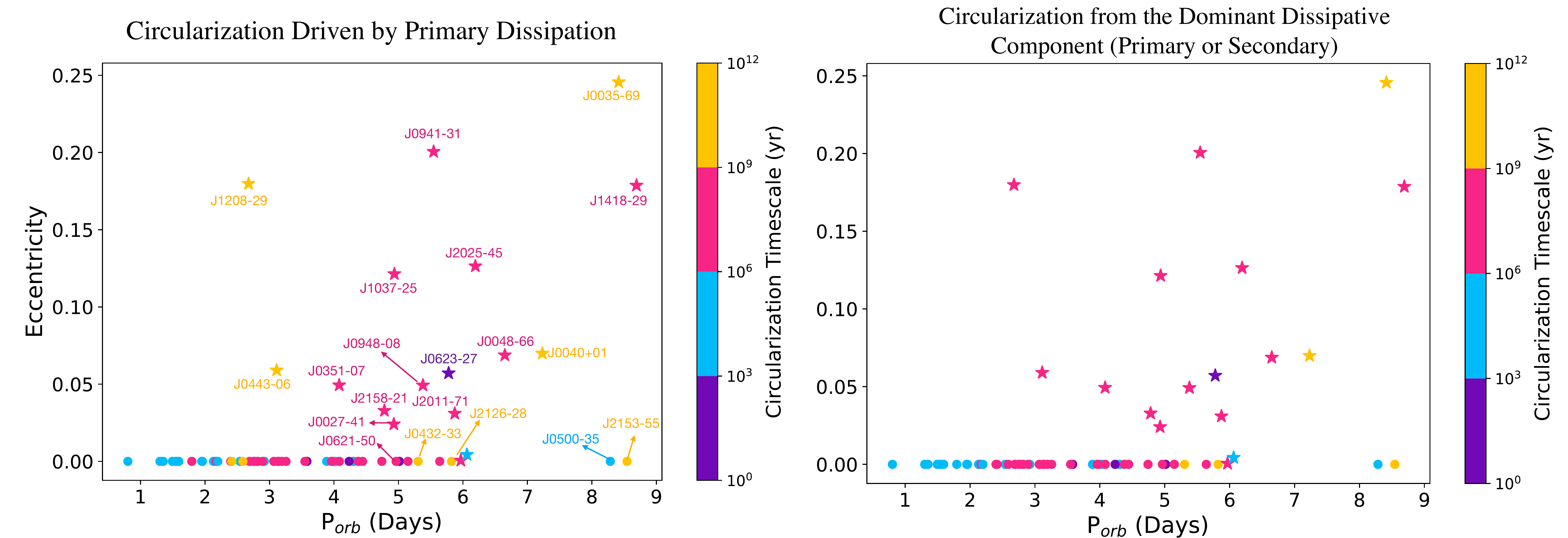}
    \caption{Eccentricity distribution of systems in our sample. Eccentricities are taken from \citet{2017A&A...EBLM4}, which reports values with a median precision of 0.0025. Circular systems are shown as circles, and eccentric systems as star symbols. \textbf{Left:} Marker colors indicate the tidal circularization timescales derived in this study due to dissipation in the primary (more massive) star.
\textbf{Right:} Marker colors indicate the minimum of the two circularization timescales due to dissipation in either the primary or the secondary star, whichever yields the shorter timescale.}
    \label{fig:circularization_timescales}
\end{figure*}

As described earlier, tides have three primary effects on the orbital evolution of a system: (1) circularization of eccentric orbits, (2) synchronization of rotational and orbital periods (aka tidal locking), and (3) spin-orbit alignment (aka decay of obliquities). In this paper, we investigate the first two of the three effects and assess whether the systems in our sample exhibit behavior consistent with what theoretical models would predict.

\subsection{Circularization Timescales} \label{sec:circ_timescales}
As the binary stars move along their eccentric orbit around a common center of mass, they experience a time-varying tidal potential. This induces a non-uniform tidal distortion over the course of the orbit, which excites time-dependent tidal flows whose kinetic energy is subsequently dissipated as heat within the stars. Over time, this process reduces the orbital eccentricity and gradually circularizes the orbit \citep{1981A&A....99..126H, 1989A&A...220..112Z}. Once the orbit becomes circular, the tidal distortion experienced by the stars remains constant throughout the orbit (as long as they are also synchronized and aligned), establishing a steady state in which no further energy is dissipated, which is when the orbit is circularized.

Based on the proof-of-principle results in \citet{2022ApJ...927L..36B}, which demonstrate that inertial wave (IW) tidal dissipation can effectively account for orbital circularization in solar-type binaries, we focus on inertial waves, which provide the dominant dissipation mechanism in our sample. For completeness, we also calculate both equilibrium tide damping by turbulent convection and dissipation of internal gravity waves under the assumption that these waves are launched from the interface between the radiative zone and convective envelope and are subsequently fully damped \citep[see][]{Baker}. These calculations are used to assess the relative importance of different dissipation channels. Inertial wave dissipation is relevant for tidal circularization if the primaries are F/G/K-type stars with convective envelopes because tidal forcing frequencies $\omega$ lie within the range of inertial waves -- if $\Omega_{\rm A}=2\pi/P_\mathrm{rot}$ is the stellar rotational angular frequency of the primary, we require $|\omega|<2\Omega_{\rm A}$, which is naturally satisfied in approximately spin-synchronized stars, since the tidal frequency of the dominant tidal component satisfies $|\omega|=\Omega_{\rm A}$.

We compute two circularization timescales for each system in our sample. The first, $\tau_{\rm e,A}$, corresponds to dissipation within the primary star (A), while $\tau_{\rm e,B}$ corresponds to dissipation within the secondary star (B). In both cases, we confirm that inertial waves provide the most efficient dissipation in all but three systems in our sample (EBLM J0345-10, J0526+04, and J0659-61), where gravity-wave dissipation is slightly more efficient. However, in these cases the resulting tidal quality factors differ by less than an order of magnitude. For consistency across the sample, we therefore adopt inertial-wave dissipation as the dominant mechanism when computing tidal timescales throughout this paper. We also assume that the stellar spins are synchronized with the orbit\footnote{This is well justified for all of the secondary stars in our sample as their estimated synchronization timescales due to inertial waves are much smaller than their ages. It is only approximately valid for the primary stars, as we shall present later, but the degree of asynchronism is insufficiently large to substantially alter these predictions.} ($P_\mathrm{rot} = P_\mathrm{orb}$) and that the orbital eccentricity $e$ is small for the purposes of these estimates. We further assume that all quadrupolar tidal components contribute equally, with the same modified tidal quality factor $Q'$, then the resulting circularization timescale is given by
\citep[e.g.][]{Baker}
\begin{equation}
    \tau_{\rm{e,A}} = \frac{2Q'}{63 \pi}\left(\frac{M_{\rm A}}{M_{\rm B}}\right)\left(\frac{M_{\rm A} + M_{\rm B}}{M_{\rm A}}\right)^{\frac{5}{3}}\frac{P_{\rm orb}^{\frac{13}{3}}}{P_{\rm dyn}^{\frac{10}{3}}},
    \label{eq:circ}
\end{equation}
where $M_{\rm A}$ and $M_{\rm B}$ are the primary and secondary star masses, respectively, and $P_{\rm dyn}$ is the dynamical timescale of star A, which is defined as 
\begin{equation} \label{P_dyn}
P_{\rm dyn} = 2\pi \sqrt{\frac{R_{\rm A}^3}{GM_{\rm A}}},
\end{equation}
where $R_{\rm A}$ is its radius. The same expression as Eq.~\ref{eq:circ} applies for $\tau_{\rm e,B}$ except that we must interchange subscripts A and B and compute Eq.~\ref{P_dyn} and $Q'$ for body B. We compute $Q'$ representing the dissipation of inertial waves according to the frequency-averaged formalism of \citet{O2013} accounting for the realistic structure of the star following \citet{Baker,2022ApJ...927L..36B} \citep[see also, e.g.,][for a similar approach adopting a piecewise homogeneous two-layer model]{Mathis2015}. To do this, we first compute stellar models for each star using Modules for Experiments in Stellar Astrophysics
\citep[MESA][]{Paxton2011, Paxton2013, Paxton2015, Paxton2018, Paxton2019} with the MIST parameters \citep{Dotter,Choi} to produce radial profiles as a function of age. 

For each binary system, we adopt the stellar age inferred using the software package {\sc bagemass} \citep{2015A&A...575A..36M} which compares the primary star's properties to a grid of stellar models computed with the {\sc garstec} stellar evolution code \citep{2008Ap&SS.316...99W}. The methods used to calculate this model grid are described in \citet{2013MNRAS.429.3645S}. {\sc bagemass} uses a Markov-chain Monte-Carlo method to explore the posterior probability distribution (PPD) for the mass and age of a star based on its observed T$_{\rm eff}$,  mean stellar density and surface metal abundance [Fe/H]. Where no direct measurements of [Fe/H] are available, we assume [Fe/H] $=0.0\pm0.3$. We evaluate Eq.~\ref{eq:circ} at the best age inferred from {\sc bagemass} where possible, but the closest stellar model fit to the inferred stellar radius was used instead for a small number of systems. Using this age-matched MESA model stellar profile, we solve for the equilibrium/non-wavelike tide by solving Eqs.~17-19 in \citet{Baker}, followed by Eqs.~28-29 of the same paper to obtain the inertial wave pressure perturbation, both done using a Chebyshev collocation method. The resulting tidal quality factor representing the dissipation of inertial waves (as it is found to be the dominant tidal dissipation mechanism in almost all cases under our assumptions, as we described above) is given by Eq.~30 of \citet{Baker}. The resulting values of $Q'$ for each star (evaluated at the stellar age inferred from {\sc bagemass}) are reported in Table \ref{Tab:Rot per}. A typical value for solar-type main-sequence stars with masses smaller than $1.1M_\odot$ is given by 
\begin{align}
    Q'\sim 10^7 \left(\frac{P_\mathrm{rot}}{10\mathrm{d}}\right)^2.
\end{align}
However, the actual numerical value can vary substantially, particularly once the stars reach the end of the main sequence \citep[see e.g.~Fig.~6 in][]{Baker}. Stars more massive than $1.1 M_\odot$ also tend to be less dissipative due to their thinner convective envelopes with a lower mean density. 

Here we caution that stellar ages inferred with {\sc bagemass} are often only moderately constrained, with typical fractional uncertainties of order $\sim$25-50\%. When propagated into the tidal timescale calculations, these age uncertainties can lead to large variations in the inferred values, particularly for the primary star (body A), whose allowed age range may extend into pre–main sequence or post–main sequence phases, where both the stellar structure and inertial-wave dissipation can vary rapidly and differ substantially to the main-sequence phase. In contrast, the secondary stars in our sample are M dwarfs that predominantly remain on the main sequence over the full allowed age range, making tidal timescales associated with dissipation inside the secondary ($\tau_{\rm e, B}$) significantly less sensitive to the age uncertainties. Systems whose tidal timescales have large uncertainties ($> 2$ dex) are marked with an asterisk in Table~\ref{Tab:Rot per} and should be interpreted with caution. The values reported in Table~\ref{Tab:Rot per} are evaluated at the best age from {\sc bagemass} and reflect the best currently-calculable tidal timescale estimates under our assumptions.

 It should also be noted that our approach is the simplest way to represent inertial wave dissipation. It ignores the tidal frequency dependence of $Q'$ that would be predicted by linear theory \citep[e.g.][]{O2013,Dewberry2024}, essentially adopting a ``typical value'' representing inertial wave dissipation throughout the whole frequency range of these waves. While this is unlikely to be valid in general -- and we should expect some outliers to theoretical expectations even if inertial waves are solely responsible for driving tidal evolution -- the frequency-averaged value is at least supported by nonlinear simulations of inertial waves \citep{AB2023}, and by its robustness \citep[e.g., by incorporation of magnetic fields,][]{LO2018}. We would expect it to typically provide shorter tidal timescales than dissipation of inertial waves in frequency-dependent calculations, so it might over-predict the efficiency of inertial wave dissipation. It is the simplest -- and currently most robust -- way to model their effects in stellar systems however, particularly for population-wide studies.

In Figure~\ref{fig:circularization_timescales} we show the observed eccentricities of binaries in our sample, color-coded by their calculated circularization timescales. 

\subsection{Synchronization Timescales}
Since stars are extended objects rather than point masses, they experience a differential gravitational force in a binary system, called the tidal force, which can distort the shape of a star from a sphere to an ellipsoid, forming a tidal bulge. When such a distorted star rotates about its spin axis while also orbiting the system's common center of mass, and the system is not synchronized (i.e., $P_{\rm orb} \ne P_{\rm rot}$), the tidal bulge lags or leads the line connecting the centers of the two bodies (the line of gravitational attraction). This misalignment generates a tidal torque that acts to realign the bulge with the line of attraction, gradually driving the system toward synchronization or tidal locking (at least for circular orbits), where $P_{\rm orb} = P_{\rm rot}$ \citep{1980A&A....Hut}. Eccentric binaries instead reach a state of pseudosynchronization, which is discussed in more detail in \S~\ref{sec:pseudosynch}
\begin{figure*}
    \centering
    \includegraphics[width = \textwidth]{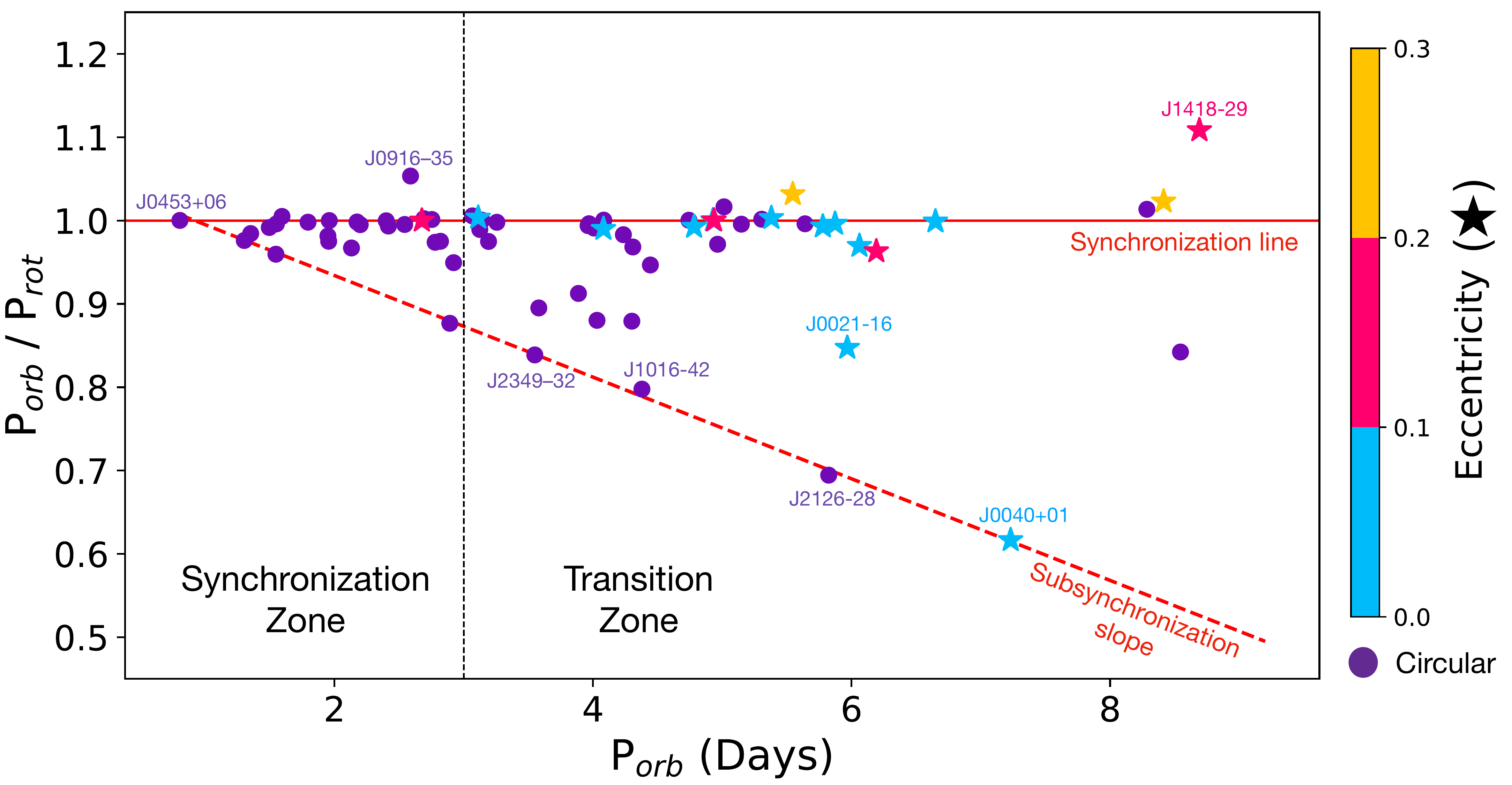}
    \caption{Period–ratio diagram for our sample. The solid red line at 
    $P_{\rm orb}/P_{\rm rot} = 1$ marks the synchronization line, with systems clustered around it identified as tidally locked. Circular EBLMs are shown as purple circles, while eccentric EBLMs are shown as star symbols, color-coded by eccentricity. Since most systems with $P_{\rm orb} \lesssim 3$ days are synchronized, this region is labeled the “Synchronization Zone”, whereas longer-period region consists of binaries showing larger deviations from synchronization and is labeled as the “Transition Zone”. An approximate best-fit straight line, representing the boundary defined by the most subsynchronous systems is shown as a dashed red line and referred to as the “subsynchronization slope”.}
    \label{fig:period_ratio}
\end{figure*}

In this paper, we examine tidal synchronization using the rotational period measurements from \citet{2024MNRAS.529.4442S}. Synchronization is achieved when the period ratio, $P_{\rm orb}/P_{\rm rot} = 1$, which is represented by the synchronization line in the period ratio diagram (Figure~\ref{fig:period_ratio}). Systems with $P_{\rm orb}/P_{\rm rot} > 1$ are called supersynchronous (since the star rotates faster than the orbit), whereas those with $P_{\rm orb}/P_{\rm rot} < 1$ are called subsynchronous (since the star rotates slower than the orbit). 

To assess whether our observations are consistent with theoretical expectations, we compute tidal synchronization timescales ($\tau_{\rm s}$) for all EBLMs in our sample. These were calculated using the same approach to that described in \S~\ref{sec:circ_timescales}, except that the synchronization timescale is given by \citep[e.g.][]{Baker}
\begin{equation}
\label{eq: synchronization timescales}
    \tau_{\rm s} = \frac{2Q' r_g^2}{9\pi}\left(\frac{M_{\rm A} + M_{\rm B}}{M_{\rm B}}\right)^2\frac{P_{\rm orb}^4}{P_{\rm dyn}^2 P_{\rm rot}}.
\end{equation}
Here $r_g^2$ is the squared radius of gyration\footnote{Defined by 
\begin{align}
r_g^2=\frac{8\pi}{3M_{\rm A}R_{\rm A}^2}\int_0^{R_{\rm A}}\rho r^4\mathrm{d}r, 
\end{align}
such that $I_{\rm A}=r_{g}^2M_{\rm A} R_{\rm A}^2$ is the moment of inertia of body A, 
where $\rho(r)$ is the stellar density profile and $r$ is the spherical radius from the stellar center. For example, $r_g^2=0.0735$ for the current Sun.} of body A and all other symbols are as defined above. We assume inertial waves are also responsible for spin synchronization and model their dissipation using the same approach as in \S~\ref{sec:circ_timescales}, with the same assumptions and caveats mentioned there. Eq.~\ref{eq: synchronization timescales} assumes a circular and aligned orbit. These timescales are evaluated at the ages inferred above. A similar calculation for the synchronization timescale for the secondary star shows that we would predict it to be spin-synchronized in all cases.

\subsection{Spin-Orbit Alignment}
Tidal dissipation in systems with spin-orbit misalignment typically acts to realign stellar spin axes with the net orbital angular momentum vector \citep[e.g.][]{BO2009,Lai2012,2015ARA&A..53..409W}. While spin-orbit misalignment has been extensively studied in the context of hot Jupiters, many of which exhibit significant misalignments, potentially indicating a dynamically active past \citep[e.g.][]{2005ApJ...631.1215W,FT2007, 2014A&A...567A..42C,2016MNRAS.456.3671A}, the phenomenon remains comparatively underexplored in stellar binaries. Among the few studies that have started to address this, \citet{1994AJ....107..306H} focused on wide binaries, while the most comprehensive measurements for close binaries come from the BANANA survey \citep{2012IAUS..282..397A,2014ApJ...785...83A}, which primarily targets massive stars. In contrast, the EBLM survey focuses on binaries with small mass ratios and solar-like primaries. 

A central objective of the EBLM collaboration has been to measure stellar obliquities in order to place constraints on tidal realignment mechanisms. To date, five true obliquity ($\psi$, the angle between the stellar spin axis and the orbital angular momentum vector) measurements have been published by the EBLM team: WASP-30 \& EBLM J1219-39 \citep{2013A&A...549A..18T}, EBLM J0218-31 \citep{2019A&A...626A.119G}, EBLM J0608-59 \citep{2020MNRAS.497.1627K}, and EBLM J0021-16 \citep{2025arXiv250921517S}. All of these systems were found to have well-aligned primaries, except for EBLM J0021-16, which shows a slight misalignment with a true obliquity of $\psi = 29.3^{\circ}$. This is also the only system included in our sample.

We may compute a tidal alignment timescale in a similar way to how we have computed spin-orbit synchronization above. In systems with more orbital than spin angular momentum, the alignment timescale is expected to be comparable to the synchronization timescale. Assuming inertial waves also dominantly contribute to driving tidal alignment \citep[e.g.][]{Lai2012,B2016,LO2017,Nils2025}, for small angles we may obtain an alignment timescale $\tau_\psi$ by computing
\begin{align}
    \tau_\psi=\frac{4}{9}\frac{Q'r_g^2}{\pi}\left(\frac{M_{\rm A}+M_{\rm B}}{M_{\rm B}}\right)^2\frac{P_\mathrm{orb}^4}{P_\mathrm{dyn}^2P_\mathrm{rot}}=2\tau_{\rm s},
\end{align}
where the latter equality is valid only if $Q'$ is the same for synchronization and alignment. Hence, our calculations of $\tau_{\rm s}$ give us
a prediction for when inertial waves will cause substantial changes in spin-orbit angles (primarily towards alignment, though evolution towards anti-alignment, or perpendicularity, are possible in principle depending on the tidal mechanism e.g.~\citealt{Lai2012,2014ARA&A..52..171O,B2016}). For these purposes, we may adopt the same $Q'$ computed above for simplicity, which is approximately justified because it gives similar values to those computed by \citet{LO2017} for obliquity tides in idealized models, even if this is for a different tidal component. 

We leave a broader observational study of spin-orbit alignment in tight binaries to a later work, once more eclipsing binary Rossiter-McLaughlin measurements have been published, from EBLM and elsewhere.

\section{Results \& Discussion} \label{sec:results}
Building on the analysis presented above, we now interpret the observed trends, their dependence on key system parameters, and examine notable outliers which deviate from theoretical expectations in the context of current tidal models. Throughout the following analysis, tidal timescales are evaluated at the best age inferred from {\sc bagemass}. While formal uncertainties in these timescales for many systems—particularly due to dissipation in the primaries— may span multiple orders of magnitude due to age uncertainties, we adopt the best age as a working reference point for comparing tidal timescales to system ages. Our interpretation is therefore aimed at understanding the broader physical picture and relating the observed eccentricities and stellar rotation periods to expectations from tidal evolution, rather than determining precise tidal timescales.
\subsection{Circularization in the EBLM Sample}
In Figure \ref{fig:circularization_timescales}, we observe that eccentric binaries in our sample only begin to appear beyond $P_{\rm orb}$ of approximately 3 days, consistent with the findings of \citet{2022ApJ...929L..27Z}. Furthermore, based on the eccentricity measurements from \citet{2017A&A...EBLM4}, 75\% of our sample appears circularized, while the remaining systems exhibit only modest eccentricities ($e<0.25$). This result is consistent with the findings of \citet{2025ApJ...982L..34W}, who showed that short-period binaries with orbital periods less than $\sim 20$ days undergo significant tidal dissipation, typically resulting in low eccentricities, similar to our observations.

To further assess whether the observed eccentricities are consistent with theoretical predictions, we look at the circularization timescales calculated in \S~\ref{sec:circ_timescales} and the inferred stellar ages. We expect that all systems with circularization timescale $\lesssim 10^9$ years have circular orbits. Contrary to this expectation, when considering dissipation only within the primary star (i.e. looking at $\tau_{\rm e, A}$ values), a subset of EBLMs are still eccentric (highlighted in Figure \ref{fig:circularization_timescales}, left)—such as EBLM J0027-41, J0048-66, J0351-07, J0623-27, J0941-31, J0948-08, J1037-25, J1418-29, J2011-71, J2025-45, J2158-21 etc. However, in many systems, dissipation within the secondary star can be equally or even more efficient. When we take the minimum of the two circularization timescales (Figure \ref{fig:circularization_timescales}, right), the discrepancy becomes even more striking: two additional systems in our sample-EBLM J1208-29 and J0443-06 have $\min(\tau_{\rm e,A},\tau_{\rm e,B}) \lesssim 1$~Gyr, implying that an even larger fraction of the sample should have circularized by now when dissipation in the secondary star is taken into account. While part of this discrepancy may stem from uncertainties in stellar age estimates or timescale calculations, another contributing factor could be that our frequency-averaged measure of inertial wave dissipation (i.e., our adopted $Q'$) could over-predict the dissipation compared with frequency-dependent calculations that follow the evolution of the system. Another explanation could be the presence of undetected tertiary companions, which can induce eccentricity oscillations through secular perturbations such as von-Zeipel-Kozai-Lidov cycles, thereby maintaining the observed eccentricities despite efficient tidal damping \citep{2016ComAC...3....6T}. Whatever the reason, this inconsistency merits further investigation.

Systems circularization timescale $\gtrsim$ Gyr are expected to still be undergoing tidal circularization and could therefore be eccentric. Considering dissipation within the primaries, these correspond to the yellow points in Figure~\ref{fig:circularization_timescales}, left. Among them, four systems— EBLM J0040+01, J0035–69, J0443-06, and J128-29, exhibit measurable eccentricities. However, when taking the minimum of the two circularization timescales, $\mathrm{min}\{\tau_{\rm e,A}, \; \tau_{\rm e,B}\}$, and accounting for dissipation in the secondary star we can confidently explain the eccentricities of only two systems, EBLM J0035-69 and EBLM J0040+01, as both exhibit circularization timescales longer than their inferred ages. Our models cannot explain the non-zero eccentricities of the remaining systems (Figure \ref{fig:circularization_timescales}, right).

Additionally, a few other systems—EBLM J0432-33, J0610-52, J0916–35, J2126–28, and J2153–55- appear circularized in our analysis despite having circularization timescales due to primary star dissipation ($\tau_{\rm e, a}$) exceeding their estimated stellar ages. Notably, three of these systems (EBLM J0432-33, J2126-28, and J2153-55) even have $\mathrm{min}\{\tau_{\rm e,A}, \; \tau_{\rm e,B}\} \gtrsim$ 1 Gyr, meaning they should not have circularized even if we account for secondary-star dissipation wherever it is the dominant component according to our models. These systems therefore represent particularly intriguing cases. While short-period binaries are generally expected to form with non-zero primordial eccentricities \citep[e.g.][]{2002MNRAS.336..705B,FT2007}, early eccentricity damping through alternative mechanisms, such as disk–binary interactions \citep[e.g.][]{2009A&A...508.1493M}, could have driven these orbits to circularity unusually early. Determining whether such systems were born circular or were circularized shortly after formation offers important clues about the formation and early evolution of close binaries. This is an open question, which we leave to future studies. A final aspect is that coupled evolution of the orbit and stellar rotation can lead to periods in which eccentricity is excited for sufficiently fast rotators \citep[e.g.][]{1981A&A....99..126H,1989A&A...223..112Z}, which may complicate this picture.

\subsection{Synchronization in the EBLM Sample} \label{results}
Based on a simple analogy with the Earth-Moon system, where the Moon is locked to Earth but not vice versa, we might expect only the secondary star to become tidally locked in unequal-mass eclipsing binaries (such as those here with $q<0.6$). Although the Earth–Moon mass ratio ($q \approx 0.012$) is far smaller than any of the binaries in our sample, this analogy provides a physical intuition. However, our results indicate otherwise. We find that approximately 78\% of our sample systems have $0.95 < P_{\rm orb}/P_{\rm rot} < 1.05$, strongly indicating that tidal forces from the M-dwarf secondaries have successfully synchronized the rotation of the surface layers of the primary stars in our sample. In the following section, we investigate how synchronization correlates with key system parameters, including orbital period, mass ratio, and primary mass.

\subsubsection{Physical Dependencies}

\begin{figure*}
    \centering
    \includegraphics[width = 0.65\linewidth]{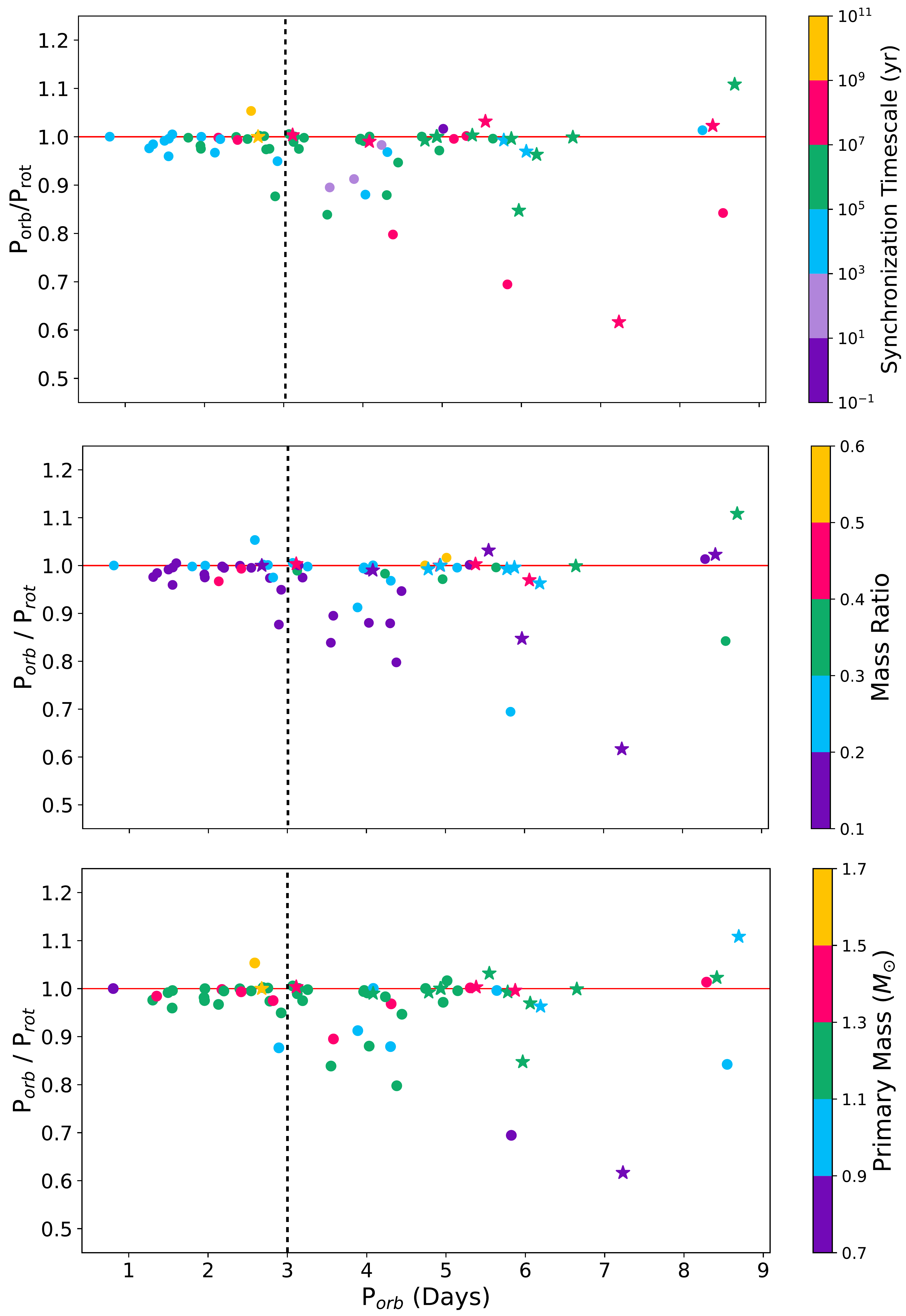}
    \caption{Dependence of the degree of synchronization on key system parameters. All three panels show period–ratio diagrams ($P_{\rm orb}/P{\rm rot}$ vs. $P_{\rm orb}$): (top) color-coded by synchronization timescale, (middle) by mass ratio ($q$), and (bottom) by primary mass ($M_{\rm A}$).}
    \label{fig:main_plot}
\end{figure*}

\begin{enumerate}
    \item \textbf{Dependence on Orbital Period : }
    Tidal synchronization is theoretically expected to depend strongly on the orbital period, with models typically predicting that synchronization timescales scale as the fourth power of the orbital period, at least if $Q'$ is independent of $P_\mathrm{orb}$ \citep{1977A&A....Zahn, Baker} as it is in Eq.~\ref{eq: synchronization timescales}. To observationally test these predictions, we examine tidal synchronization within our sample. The period ratio diagram, Figure~\ref{fig:period_ratio}, reveals that nearly all systems with orbital periods shorter than $\sim 3$ days cluster tightly around the synchronization line, indicating that they are tidally locked. We refer to this region as the Synchronization Zone. For systems with orbital periods longer than 3 days, we find that approximately 26\% of the sample exhibit subsynchronous rotation, with $P_{\rm orb} / P_{\rm rot} < 0.95$. Nevertheless, the majority, about 72\% remain consistent with synchronization, lying within the range $0.95 < P_{\rm orb} / P_{\rm rot} < 1.05$. Figure \ref{fig:period_ratio} also reveals a peculiar trend marking the transition from synchronous to subsynchronous systems. As expected, the deviation from synchronization increases with orbital period; however, we find that the binaries defining the lower boundary—corresponding to the most subsynchronous systems (i.e., the maximum deviation from synchronization) lie almost perfectly along a straight line with a negative slope in the period ratio diagram. Fitting a straight line to this boundary yields the following relation:
    \begin{equation} \label{eq: sethi slope}
        \frac{P_{\rm orb}}{P_{\rm rot}} = mP_{\rm orb} + c
    \end{equation}
    where $m = -0.06098 \pm 0.00186$ is the best-fit slope and $c = 1.05604 \pm 0.00936$. We call this linear boundary, the “\textit{subsynchronization slope}".
    
    \hspace{\parindent} To test whether this slope could arise by chance, we performed 1000 Monte Carlo realizations of a lower-envelope test. We first assume $P_{\rm orb}/P_{\rm rot} = 1 - \delta$ with $\delta > 0$ for subsynchronous systems. From the observed sample, we compute the distribution of $\delta$ and estimate its standard deviation ($\sigma_{\rm obs}$). For each realization, we generate a synthetic dataset by drawing an independent random $\delta_i$ value for every observed system from a truncated normal distribution centered at 0 (representing predominantly synchronized systems) with a standard deviation of $3\sigma_{\rm obs}$, restricted to $\delta_i > 0$. Each set of draws thus produces a complete mock dataset with corresponding rotation periods ($P_{\rm rot, i}^{\rm mock} = P_{\rm orb, i}/(1-\delta_i)$), using the observed $P_{\rm orb}$ values. For each realization, we then fit a 5th-percentile (lower-envelope) quantile regression line such that approximately 95 \% of the points lie above it, thereby defining a “null” subsynchronization slope corresponding to randomly scattered $P_{\rm rot}^{\rm mock}$ values. Finally, we compare all 1000 null slopes from each realization with the observed subsynchronization slope and identify those for which at least 50\% of the null slope line lies within the 5$\sigma$ uncertainty region of the observed relation. Fewer than 1\% of the realizations satisfy this criterion, indicating that the observed subsynchronization slope is highly unlikely to arise by random chance.
    
    \hspace{\parindent} The subsynchronization slope may point to a common underlying physical mechanism that regulates the maximum degree of subsynchronization attainable at a given $P_{\rm orb}$. Possible explanations could be a different scaling of the tidal torque and corresponding timescales with rotation rate than predicted by Eq.~\ref{eq: synchronization timescales}, enhanced surface differential rotation in slower rotators (see \S~\ref{sec:dev from synch}), or the influence of variations in magnetic braking efficiencies. The fact that these systems fall along a single relation suggests that the interplay between tidal dissipation and angular momentum loss processes may follow a consistent trend across our sample, providing a potentially valuable constraint for future theoretical models. We emphasize that this trend is based on only 5 systems. Comparing with the results from \citet{2025arXiv250104082H}, which contains a larger sample that covers all binary mass ratios (not just EBLM-like), the majority of asynchronous systems do appear above Eq.~\ref{eq: sethi slope}, with a few scattered outliers below (see their Figures 11 and 12). It remains to be seen what is the cause of their anomalously slow rotators at short orbital periods.  A future more detailed exploration of the robustness of this trend is encouraged.
\\
    \item \textbf{Dependence on Mass Ratio : }
    The mass ratio, $q$ ranges from 0 to 1, with $q \rightarrow 0$ for highly unequal mass binaries, such as those in our sample, and $q = 1$ for equal mass binaries. Figure~\ref{fig:main_plot} (middle panel) illustrates the relationship between tidal synchronization and mass ratio across our sample. We observe that all binaries, even the ones with the lowest mass ratios ($0.1 < q < 0.2$) are synchronized within the synchronization zone (i.e., for $P_{\rm orb} \lesssim 3$ days). Beyond the 3-day orbital period, binaries with very low mass ratios show the most significant deviations from synchronization. In fact, the most subsynchronous binary in our sample, J0040+01 has a mass ratio of just $\sim 0.126$, $P_{\rm rot} \sim 11.73$ days, and $P_{\rm orb}$ of $\sim 7.23$ days, lying farthest from the synchronization line. 
    EBLMs with mass ratios in the range $0.2 < q < 0.3$ remain mostly synchronized up to an orbital period of $\sim 6$ days, near which, J2126–28 shows noticeable deviation from synchronization. Systems with higher mass ratios (q $\gtrsim$ 0.3) tend to stay close to the synchronization line at even longer periods, suggesting that higher mass-ratio binaries are more likely to maintain synchronization.
    While orbital period is the dominant factor, these trends demonstrate that mass ratio also plays a key role in tidal synchronization, consistent with theoretical predictions (as seen in Eq. \ref{eq: synchronization timescales}) and previous observational studies \citep{2017AJ....Lurie}. Although the overall trend is evident, a more rigorous quantitative analysis, such as identifying precise synchronization thresholds across mass ratio bins, will require a larger sample.
\\
    \item \textbf{Dependence on Primary Mass : }
    To test the dependence of synchronization on primary mass, we examine the period ratio diagram color-coded by primary mass (Figure~\ref{fig:main_plot}, bottom panel). Two of the three least massive primaries in our sample ($0.7 M_\odot < M_{\rm A} < 0.9 M_\odot$)— J2126–28 and J0040+01, are among the most subsynchronous systems and lie farthest from the synchronization line. However, J0453+06, which has a similarly low primary mass, is synchronized, likely due to its very short orbital period of just $\sim 0.804$ days. This suggests that while low primary mass may contribute to delayed synchronization, the trend is not consistent; short-period systems can still synchronize regardless of primary mass in our sample. Additionally, for stars with $M_{\rm A} > 0.9 M_\odot$, we observe a wide range of synchronization behavior. Some are synchronized, while others, especially those with very low mass ratios remain subsynchronous. Overall, there is no strong or systematic dependence on primary mass alone.
    From our analysis, it appears that mass ratio has a more significant impact, with orbital period remaining the dominant factor when it comes to synchronization. This is consistent with the findings of \citet{2017AJ....Lurie}, who also concluded that primary mass is not a strong predictor of synchronization among F/G/K-type stars.
\end{enumerate}

\subsection{Deviations from Synchronization} \label{sec:dev from synch}

Based on the estimated ages and synchronization timescales calculated for the binaries in our sample, all those having $\tau_{\rm s} <10^9$ yrs (i.e. all systems in our sample except EBLM J0916-35, and J1208-29) are expected to have reached synchronization by now. However, our period ratio diagram reveals several subsynchronous (e.g., J0040+01, J2126–28, J1016-42, etc.) and even supersynchronous systems (e.g., J1418–29) that deviate from this expectation. While these may be genuine outliers and thus compelling targets for future study, there are also plausible physical mechanisms that could account for their apparent asynchronous rotation. Two leading explanations are pseudo-synchronization and differential rotation. 

\begin{figure}
    \centering
    \includegraphics[width=0.5\textwidth]{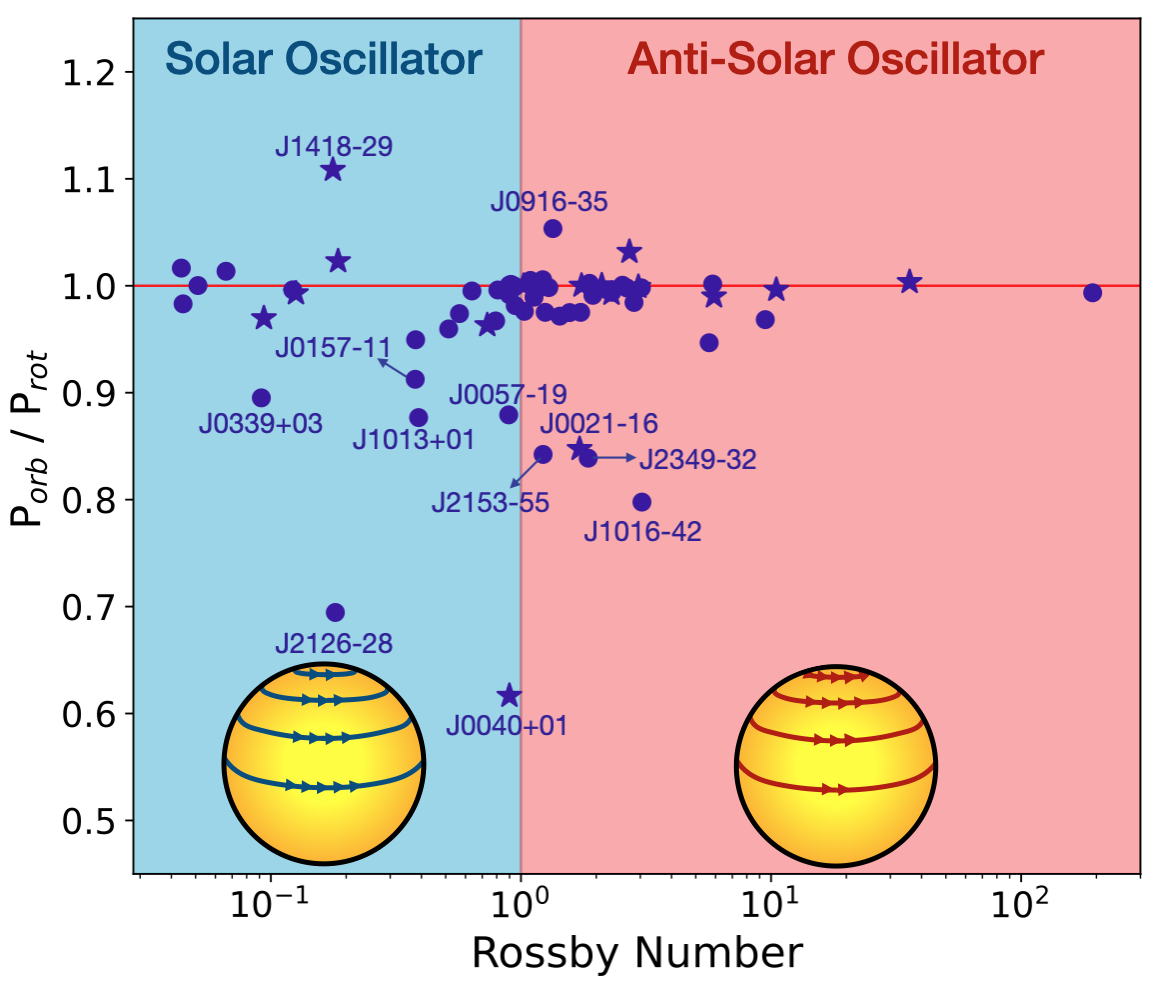}
    \caption{Convective Rossby number ($\mathrm{Ro}_c$) distribution for the systems in our sample. The blue-shaded region corresponds to $\mathrm{Ro}_c < 1$ and the red-shaded region to $\mathrm{Ro}_c > 1$. Eccentric binaries are denoted by stars, like in other plots. Binaries with the primary star in the blue region are expected to exhibit primarily solar-type differential rotation (faster at the equator), whereas those in the red region are expected to primarily exhibit antisolar-type differential rotation (faster at the poles).}
    \label{fig:rossby_no}
\end{figure}
\subsubsection{Pseudo-synchronization} \label{sec:pseudosynch}
It is often thought that eccentric binaries cannot fully synchronize and become tidally locked, but they instead become pseudosynchronized, where the rotational angular velocity becomes a significant fraction of the orbital angular velocity at periastron \citep{1981A&A....99..126H,1986ApJ...309L..83H}. Such pseudosynchronization would predict, for small $e$,
    \begin{equation} \label{eq: pseudo}
        \frac{P_\mathrm{orb}}{P_\mathrm{rot}}\approx 1+6e^2+\frac{3}{8}e^4+O(e^6),
    \end{equation}
    according to \citet{1981A&A....99..126H}, and hence stars with larger $e$ would rotate comparatively faster than the synchronized value if they were pseudosynchronized in this way. It remains unclear whether such classical pseudosynchronization would be expected in reality, however, since this was derived based on a constant time-lag tidal model, which is only approximately valid to describe equilibrium tide damping for very low tidal frequencies \citep[e.g.][]{IvPap2004,DBJ2020,VB2020,dVBH2023,Barker2025}. It is unlikely to be valid for inertial wave dissipation \citep[e.g.][]{Dewberry2024}, which is likely to be much more effective. This issue remains an important topic to study theoretically.
    
   Our results enable us to explore whether or not such pseudosynchronization is achieved. If it is, since angular velocity at periastron is higher than the mean orbital angular velocity, pseudosynchronized systems might appear supersynchronous in our analysis. For example, J1418–29 appears supersynchronous, with $e = 0.1787\pm0.0049$. Pseudosynchronization, would predict $P_\mathrm{orb}/P_\mathrm{rot}\approx 1.19$ (using Eq.~\ref{eq: pseudo}) for this system (compared with the value of approximately 1.1 observed), providing a plausible explanation for its supersynchronous state.
    An intriguing result, however, is that several other eccentric EBLMs in our sample lie precisely on the synchronization line, whereas they should appear supersynchronous if they were pseudosynchronized according to the above formula (Eq. \ref{eq: pseudo}). This behavior is discussed in more detail in the following section on differential rotation.
    
\subsubsection{Differential Rotation} In a differentially rotating star, different latitudes and potentially different radii rotate with different angular velocities \citep{2005MNRAS.357L...1B,2016MNRAS.461..497B}. Tidal synchronization is expected to occur predominantly in stellar regions where tidal flows are dissipated and tidal torques act, which may or may not coincide with the equator or even the stellar surface\footnote{For example, \citet{AB2022} found in some of their idealized simulations of tidally-driven inertial waves that synchronization occurred first in the polar regions, although it remains to be seen what would happen in more realistic models.}\citep[e.g.][]{ 1977A&A....Zahn,2014ARA&A..52..171O,AB2022,Barker2025}. In addition, even if the mean stellar rotation (in radius and latitude) is synchronized, if differential rotation is maintained by other processes such as turbulent convection, it is not clear which latitudes at the surface would be expected to indicate this. For example, the rotation rate of the solar radiative zone is the same as that at mid-latitudes in the convection zone, so if such a star is synchronized in a mean sense, this may instead be indicated by mid-latitudes rather than the equator. Nevertheless, assuming that synchronization occurs at the equator is a reasonable starting point for aligned systems. Under this assumption, if starspot overdensities are located at higher latitudes, the measured rotation period may differ from the true equatorial value, leading to apparent asynchronous systems in our analysis. In solar-type differential rotation, where the equator rotates faster than the poles, spot modulation from high latitudes yields longer measured rotation periods than the equatorial value. As a result, such systems can appear slightly subsynchronous in the period-ratio diagram (Figure~\ref{fig:period_ratio}), even if they are synchronized at the equator. On the other hand, in antisolar-type differential rotation, where the poles rotate faster than the equator, spot modulation from higher latitudes produces shorter measured rotation periods, making synchronized systems appear supersynchronous in our analysis. If instead a mid-latitude is (or the polar regions are) predominantly synchronized, this interpretation would differ accordingly.

To predict whether a system is likely to exhibit solar-type or antisolar-type differential rotation, we calculate the convective Rossby number ($\mathrm{Ro}_c$) for all binaries in our sample. $\mathrm{Ro}_c$ quantifies the relative importance of inertial forces to Coriolis forces in a rotating convective system. This was calculated using the same stellar models for the tidal computations, based on the convective velocities $u_c$ and mixing-lengths $l_c$ (as predicted by {\sc MESA}) according to $\mathrm{Ro}_c=u_c/(l_c\Omega_{\rm A})$. The radius chosen to evaluate this was one pressure scale height above the base of the convection zone of body A (more massive star). It should be remembered that this quantity is not a single number for any star, and so the rotational influence on convection will vary with radius, in general. Since the surface layers are the ones observed, where convective velocities are typically larger, and convective length-scales smaller, the effective Rossby number in surface layers could potentially be much larger.

Previous numerical simulations of rotating, convective turbulence in spherical shells, suggest that, for $\mathrm{Ro}_c \lesssim 1$, stars are expected to exhibit solar-type differential rotation. This means that the equator rotates faster than the poles. In contrast, for $\mathrm{Ro}_c \gtrsim 1$, these simulations suggest that stars are likely to have antisolar differential rotation, i.e.~for the poles to rotate faster than the equator \citep[e.g.][]{Gastine2013,2014MNRAS.438L..76G,Brun2017,Noraz2022}. Note that this interpretation can be modified by additional effects, such as magnetic fields \citep[e.g.][]{Hotta2022}, or by simulating sufficiently turbulent convection \citep[e.g.][]{Noraz2025}, so this transition at $\mathrm{Ro}_c\approx 1$ may not be definitive for interpreting observations.

With these caveats in mind, the distribution of $\mathrm{Ro}_c$ in our sample is shown in Figure \ref{fig:rossby_no}. Of the two clearly supersynchronous systems in our sample, J0916–35 has a circular orbit but exhibits $\mathrm{Ro}_c > 1$, consistent with antisolar-type differential rotation, which is therefore a plausible explanation for its observed supersynchronous rotation. Its estimated synchronization timescale further suggests that the system may not yet have fully synchronized and could therefore become even more supersynchronous over time if the latitudinal differential rotation produced by convection persists. In contrast, the calculated $\tau_{\rm s}$ of J1418–29 indicates that it should already have synchronized. We also find that it has $\mathrm{Ro}_c <1$, indicative of solar-type differential rotation, which makes differential rotation an unlikely interpretation for explaining its supersynchronous state assuming that synchronization occurs at the equator and spots are preferentially located at higher latitudes. As described above, the most plausible explanation for J1418–29’s behavior remains pseudosynchronization, even if the precise value that we would expect theoretically is presently uncertain. 

As noted earlier, several eccentric EBLMs in our sample lie on the synchronization line despite the expectation that they should appear supersynchronous if pseudosynchronized. One possible explanation is solar-type differential rotation, which could cause them to appear slightly subsynchronous and thus closer to the synchronization line. However, Figure~\ref{fig:rossby_no} shows that most of these apparently synchronized eccentric systems have $\mathrm{Ro}_c > 1$, indicative of antisolar-type differential rotation, which would instead make them appear even more supersynchronous rather than synchronized according to the above interpretation. The fact that these systems appear perfectly synchronized despite their eccentricity warrants further investigation, which is beyond the scope of this work. We note, however, that spin evolution on eccentric orbits is uncertain theoretically \citep[e.g.][]{Dewberry2024}.

Solar-type differential rotation may similarly explain the subsynchronous population we observe in the $\sim 3$-10 day orbital period range (e.g., J0157–11, J0339+03, J1013+01, J0057-19, J2126-28, and J0040+01). Moreover, as seen in Figure \ref{fig:period_ratio}, the degree of subsynchronous rotation appears to increase with the rotation period. This trend is consistent with previous studies showing that differential rotation is stronger in slower rotators than in rapid ones \citep{2003A&A...412..813R, 2007AN....328.1030C}. However, a few subsynchronous systems in our sample, particularly J2153–55, J0106–42, J2349-32, and J0021–16, have $\mathrm{Ro}_c > 1$, suggesting that they likely exhibit antisolar-type differential rotation, which is inconsistent with the subsynchronous rotation we observe. While both of the above-mentioned discrepancies could arise from our initial assumption that synchronization occurs at the equator and if spot overdensities are located at higher latitudes, one could explain the apparent deviations from synchronization observed in some systems in our sample, they may also indicate that these systems are not yet synchronized, despite synchronization timescales suggesting they should be. Such systems therefore represent valuable targets for future studies aimed at refining tidal synchronization models.

\subsection{Notes on Selected Targets}

Here we summarize the characteristics of specific targets that exhibit particularly interesting observational features, as discussed in detail throughout this paper. 

\begin{itemize}
   \item \textbf{EBLM J0021-16 : } Has a measured Rossiter-McLaughlin true 3D obliquity, showing slight misalignment between the primary star's spin axis and the orbital angular momentum vector of the binary system, $\psi=29.3\pm2.1^{\circ}$ \citep{2025arXiv250921517S}. Its predicted $\tau_{\rm s}$ is much shorter than the estimated system age, indicating that it should be synchronized; however, it is observed to be subsynchronous. With $\mathrm{Ro}_c >1$, the system is likely to exhibit antisolar-type differential rotation, which may not be able to account for the observed subsynchronous rotation. The system also exhibits a much shorter circularization timescale ($\min(\tau_{\rm e,A},\tau_{\rm e,B}) \approx 84$ Myr) than its predicted age, implying that it should have circularized; yet it still retains a small eccentricity of 0.000550 ± 0.000090. This makes it an intriguing target for future studies.

   \item \textbf{EBLM J0027-41, J0048-66, J0345-10, J0351-07, J0623-27, J0941-31, J0948-08, J1037-25, J1418-29, J2011-71, J2025-45, J2158-21 : } Binaries with predicted circularization timescales—both $\tau_{\rm e,A}$ and $\tau_{\rm e,B}$, corresponding to dissipation within the primary and secondary stars, respectively—shorter than their estimated stellar ages, yet exhibiting small but measurable eccentricities.
   
    \item \textbf{EBLM J0916-35 : } A circular binary ($M_{\rm A} = 1.61 M_\odot, M_{\rm B} = 0.349 M_\odot$) It has $P_{\rm orb} \approx 2.59$ days and a supersynchronously rotating primary star having $P_{\rm rot} \approx 2.46$ day. Since the orbit is circular, the system cannot be pseudosynchronized; however, $\mathrm{Ro}_{c} > 1$ suggests an antisolar-type differential rotation as a plausible explanation for this behavior.
    \item \textbf{EBLM J1418-29 : } Binary with $M_{\rm A} = 1.03 M_\odot, M_{\rm B} = 0.327 M_\odot$, and an eccentric orbit ($e \approx 0.18$) on a $\sim 8.69$ day $P_{\rm orb}$.
    The primary is supersynchronous with a measured $P_{\rm rot} \approx 7.84$ days, likely due to pseudosynchronization.
    
    \item \textbf{EBLM J0040+01,  EBLM J0057-19, EBLM J0157-11, EBLM J0339+03,, EBLM J2126-28 : } Binaries with predicted $\tau_{\rm s}$ shorter than the estimated system ages and therefore would be expected to be synchronized; however, they are observed to be subsynchronous. For all of them, $\mathrm{Ro}_c < 1$, suggesting that the slight subsynchronous behavior may be attributed to solar-type differential rotation. 
    \item \textbf{EBLM J1016-42, EBLM J2153-55, EBLM J2349-32 : } Binaries with predicted $\tau_{\rm s}$ shorter than their estimated ages and hence we would expect them to be synchronized; however, these systems are observed to be subsynchronous. All have $\mathrm{Ro}_c >1$, indicative of antisolar-type differential rotation, which is less easily able to account for their subsynchronous behavior. They therefore represent intriguing targets for future studies.
    \item \textbf{EBLM J0432–33, J2126–28, and J2153–55}: Systems having $\min(\tau_{\rm e,A},\tau_{\rm e,B}) \gtrsim 1$ Gyr, indicating that they shouldn't have circularized given their inferred ages, yet they are observed to have circular orbits. They represent intriguing cases for understanding the early formation and evolutionary pathways of short period binary systems.
\end{itemize}

\section{Conclusion} \label{sec:conclusion}
We investigate tidal effects—specifically circularization and synchronization—in 68 unequal mass ($0.1 \le q \le 0.6$), short-period ($P_{\rm orb} < 10$ days) F/G/K + M eclipsing binaries from the EBLM survey. A summary of key results is as follows:

\begin{enumerate}[label=\arabic*.]
    \item Approximately 75\% of our sample is circularized. Eccentric binaries begin to appear beyond an orbital periods of 3 days; however, the sample still shows only small eccentricities ($e < 0.25$). 
    
    \item A few binaries have theoretical circularization timescales (based on inertial waves in convective envelopes which we find to be the dominant dissipative mechanism in our sample, following \citealt{Baker,2022ApJ...927L..36B}) shorter than their estimated system ages, yet they remain non-circular. This discrepancy could point to the influence of a tertiary companion, which future observations may be able to detect. Alternatively, they could require more sophisticated tidal models than those adopted here. 
    
    \item A few systems in our sample appear circular despite having theoretical circularization timescales due to dissipation in the dominant component of the binary ($\min(\tau_{\rm e,A},\tau_{\rm e,B}$) longer than their estimated ages. One possibility is that these binaries were “born” circular (or nearly so). Alternatively, early eccentricity damping through mechanisms such as circumbinary disk interactions could have circularized the orbits shortly after formation. It may also suggest limitations in our current tidal models, which may not fully capture the relevant dissipation processes at play.
    
    \item Using rotation periods from \citet{2024MNRAS.529.4442S}, we find that 78\% of our sample is tidally locked, demonstrating that even a low-mass M-dwarf can synchronize a solar-type primary in a tight binary. We find a strong correlation between tidal locking and orbital period. Nearly all systems having $P_{\rm orb} < 3$ days are synchronized, defining the synchronization zone. Beyond this region, most binaries ($\sim 72$\%) remain synchronized \citep{2017AJ....Lurie}, though a few exhibit subsynchronous or supersynchronous rotation. Overall, our analysis shows that while orbital period has the strongest influence on the degree of synchronization, it is also shaped by mass ratio, whose impact is more significant than that of the primary mass alone.

    \item A clear trend marks the transition from synchronous to subsynchronous systems. The binaries that define the lower boundary (i.e. the most subsynchronous systems) lie almost perfectly along a straight line with a negative slope in the period–ratio diagram. This well-defined linear boundary referred to as the “subsynchronization slope” in the paper, may indicate that there is an underlying physical mechanism regulating the maximum degree of subsynchronization attainable at a given orbital period from the interaction of tidal synchronization processes and the mechanisms driving differential rotation. Such a trend carries important implications for theoretical studies of tidal synchronization and may help to constrain the physical processes that govern it. 

    \item We explored plausible explanations for the asynchronous systems in our sample. One is pseudosynchronization, where eccentric binaries cannot fully synchronize but instead rotate near the orbital angular velocity at periastron and thus appear supersynchronous (e.g., EBLM J1418–29). However, several eccentric systems lie directly on the synchronization line, contrary to this expectation. Differential rotation may offer a partial explanation: solar-type profiles can make synchronized systems appear subsynchronous, while antisolar-type profiles can make them appear supersynchronous (at least if tides preferentially synchronize the equator and spot modulations measure rotation at higher latitudes). The convective Rossby numbers of these stars can be used to infer whether solar-like (when this is smaller than one) or anti-solar-like (when it exceeds one) differential rotation would be expected. For our stars, they provide evidence for both regimes, though not all systems fit cleanly within this framework. Some subsynchronous binaries in the 3–10 day range are consistent with solar-type differential rotation, but others remain puzzling, with $\mathrm{Ro}_c$ values inconsistent with their observed states. These discrepancies may reflect the limitations of assuming equatorial synchronization or indicate systems that are not yet synchronized, making them valuable targets for refining tidal synchronization models.
\end{enumerate}

By examining this EBLM sample of unequal mass binaries, we probe the limits of tidal processes, identifying where tides are most efficient in driving circularization and synchronization, and where their influence weakens. It would be beneficial in future work to expand the sample of binaries, particularly by including systems with orbital periods longer than those considered here. A larger and more diverse sample would provide stronger constraints on theoretical models and help confirm the population-level trends identified in this study, such as the observed subsynchronization slope. Additionally, since our interpretations are conditioned on stellar ages inferred from {\sc bagemass} and the corresponding tidal timescales evaluated at the returned best age; future improvements in age constraints would further refine and strengthen the interpretations presented here.

\section*{Data availability}
All data used in this analysis (e.g. light curves) can be provided upon reasonable request to the authors.

\section*{Acknowledgements}

Support for this work was provided by NASA through the NASA Hubble Fellowship grant HF2-51464 awarded by the Space Telescope Science Institute, which is operated by the Association of Universities for Research in Astronomy, Inc., for NASA, under contract NAS5-26555. This work was based on photometry taken by the TESS spacecraft. In particular, we benefited from short cadence 120 second photometry, courtesy of Guest Investigator programs led by the EBLM team. AJB was supported by STFC grants ST/W000873/1 and UKRI1179. AHMJT is supported by the ERC/UKRI Frontier Research Guarantee programme (EP/Z000327/1/CandY). Part of this research from AD was carried out at the Jet Propulsion Laboratory, California Institute of Technology, under a contract with the National Aeronautics and Space Administration (80NM0018D0004).

 \bibliographystyle{mnras}
 \bibliography{references}
 \begin{table*}
     \centering
     \caption{Values marked with an '$*$' indicate high uncertainty, for which the propagated tidal timescale uncertainty exceeds 2 dex. These large uncertainties arise from poorly constrained stellar ages, whose allowed ranges span multiple evolutionary phases, including the pre–main sequence and post–main sequence (giant branch). The uncertainty in $\log_{10}\rm Age$ is typically $\pm$ 0.1–0.2. }
     \label{Tab:Rot per}
     \resizebox{1.05\linewidth}{!}{
     \begin{tabular}{|l|l|l|l|l|l|l|l|l|l|l|l|} 
     \hline
\textbf{EBLM Name} & $\mathbf{P_{\rm orb}}$ & \textbf{Nominal} $\mathbf{P_{\rm rot}}$ & \textbf{Eccentricity} & $\mathbf{M_{\rm A}}$ & $\mathbf{M_{\rm B}}$ & \textbf{q} & $\mathbf{\tau_{\rm s}}$ & $\mathbf{\tau_{\rm e, A}}$ & $\mathbf{\tau_{\rm e, B}}$ & $\mathbf{Q'}$ & $\log_{10}\left(\frac{\rm Age}{\rm Gyr}\right)$ \\ 
& \textbf{[days]}&\textbf{[days]}& &\textbf{[$\mathbf{M_\odot}$]}&\textbf{[$\mathbf{M_\odot}$]}& & $\mathbf{[yr]}$ & $\mathbf{[yr]}$ & $\mathbf{[yr]}$ &  & \\ \hline 
EBLM J0021-16 & 5.96727 & 7.04±0.06 & 0.000550 ± 0.000090 & 1.11±0.07 & 0.194±0.022 & 0.1748 & 2.99$\times 10^{6*}$ & 8.44$\times 10^{7*}$ & 7.67$\times 10^{9}$ & 1.77$\times 10^{6}$ & 0.8 \\
EBLM J0027-41 & 4.92799 & 4.93±0.16 & 0.02409±0.00115 & 1.18±0.07 & 0.528±0.065 & 0.4475 & 1.89$\times 10^6$ & 9.96$\times 10^{7}$ & 5.77$\times 10^{8}$ & 4.85$\times 10^{6}$ & 0.5 \\
EBLM J0035-69 & 8.4146 & 8.23±0.09 & 0.24562±0.00075 & 1.17±0.07 & 0.198±0.02 & 0.1692 & 2.48$\times 10^{8*}$ & 1.10$\times 10^{10*}$ & 5.69$\times 10^{10}$ & 2.45$\times 10^{5}$ & 1.0 \\
EBLM J0040+01 & 7.231 & 11.73±0.13 & 0.06988±0.00049 & 0.81±0.06 & 0.102±0.01 & 0.1259 & 3.55$\times 10^8$ & 1.78$\times 10^{10*}$ & 5.59$\times 10^{9}$ & 6.08$\times 10^{6}$ & -1.2 \\
EBLM J0048-66 & 6.64927 & 6.66±0.01 & 0.06876±0.0008 & 1.25±0.09 & 0.447±0.055 & 0.3576 & 1.73$\times 10^{6*}$ & 8.00$\times 10^{7*}$ & 8.69$\times 10^{9*}$ & 4.61$\times 10^{6}$ & 0.6 \\
EBLM J0057-19 & 4.30051 & 4.89±0.08 & $\le0.0062$ & 1.09±0.08 & 0.136±0.019 & 0.1248 & 1.58$\times 10^{6*}$ & 2.02$\times 10^{7*}$ & 3.93$\times 10^{9}$ & 1.15$\times 10^{6}$ & 0.8 \\
EBLM J0157-11 & 3.88745 & 4.26±0.11 & $\le0.00575$ & 1.07±0.08 & 0.309±0.043 & 0.2888 & 1.92 $\times 10^{-1*}$ & 1.7$\times 10^{3*}$ & 7.56$\times 10^{7}$ & 1.34$\times 10^{3}$ & 1.0 \\
EBLM J0239-20 & 2.77868 & 2.85±0.12 & $\le0.0033$ & 1.16±0.08 & 0.17±0.029 & 0.1466 & 1.29$\times 10^{5*}$ & 8.97$\times 10^{5*}$ & 1.08$\times 10^{8}$ & 4.05$\times 10^{5}$ & 0.7 \\
EBLM J0247-51 & 4.00785 & 4.04±0.06 & $\le0.00607$ & 1.26±0.11 & 0.231±0.038 & 0.1833 & 4.30$\times 10^{5*}$ & 5.48$\times 10^{6*}$ & 3.44$\times 10^{8}$ & 1.91$\times 10^{6}$ & 0.6 \\
EBLM J0326-09 & 2.4004 & 2.40±0.05 & $\le0.04123$ & 1.16±0.08 & 0.193±0.038 & 0.1664 & 2.74$\times 10^5$ & 2.29$\times 10^{6}$ & 2.68$\times 10^{7}$ & 1.11$\times 10^{6}$ & 0.5 \\
EBLM J0339+03 & 3.58067 & 4.00±0.02 & $\le0.00504$ & 1.33±0.18 & 0.241±0.051 & 0.1812 & 2.57$\times 10^{1*}$ & 2.57 $\times 10^*$ & 1.45$\times 10^{8}$ & 1.70 $\times 10$ & 0.6 \\
EBLM J0345-10 & 6.06135 & 6.25±0.03 & 0.00433±0.00039 & 1.21±0.09 & 0.525±0.066 & 0.4339 & 9.50$\times 10^{3*}$ & 8.99$\times 10^{3*}$ & 2.06$\times 10^{9}$ & 5.96$\times 10^{5}$ & 0.8 \\
EBLM J0351-07 & 4.0809 & 4.12±0.05 & 0.04936±0.00137 & 1.29±0.11 & 0.242±0.04 & 0.1876 & 1.11$\times 10^7$ & 1.11$\times 10^{7}$ & 3.62$\times 10^{8}$ & 1.82$\times 10^{7}$ & 0.3 \\
EBLM J0356-18 & 1.30211 & 1.33±0.03 & $\le0.08078$ & 1.29±0.1 & 0.134±0.043 & 0.1039 & 2.64$\times 10^4$ & 5.38$\times 10^{4*}$ & 2.31$\times 10^{6}$ & 6.21$\times 10^{5}$ & 0.4 \\
EBLM J0400-51 & 2.69208 & 2.69±0.04 & $\le0.01869$ & 1.26±0.1 & 0.266±0.052 & 0.2111 & 3.19$\times 10^{5*}$ & 3.44$\times 10^{6*}$ & 1.72$\times 10^{7}$ & 2.42$\times 10^{6}$ & 0.4 \\
EBLM J0432-33 & 5.30549 & 5.30±0.01 & $\le0.00307$ & 1.32±0.09 & 0.26±0.035 & 0.197 & 2.00$\times 10^{8*}$ & 6.21$\times 10^{9*}$ & 1.60$\times 10^{9*}$ & 1.10$\times 10^{8}$ & -0.3 \\
EBLM J0440-48 & 2.54304 & 2.56±0.06 & $\le0.01504$ & 1.29±0.1 & 0.19±0.039 & 0.1473 & 1.03$\times 10^{5*}$ & 5.83$\times 10^{5*}$ & 4.28$\times 10^{7}$ & 8.61$\times 10^{5}$ & 0.5 \\
EBLM J0443-06 & 3.11192 & 3.10±0.08 & 0.05899±0.00123 & 1.39±0.13 & 0.577±0.108 & 0.4151 & 4.24$\times 10^{7*}$ & 1.26$\times 10^{9*}$ & 4.36$\times 10^{7*}$ & 4.73$\times 10^{8}$ & -0.2 \\
EBLM J0453+06 & 0.80415 & 0.80±0.01 & $\le 0.11522$ & 0.78±0.06 & 0.17±0.049 & 0.2179 & 1.07$\times 10^3$ & 2.85$\times 10^{3}$ & 3.25$\times 10^{4}$ & 2.74$\times 10^{4}$ & 1.2 \\
EBLM J0454-09 & 5.01345 & 4.93±0.07 & $\le0.00177$ & 1.18±0.12 & 0.708±0.104 & 0.6 & 5.69 $\times 10^{0*}$ & 1.11 $\times 10$ & 1.53$\times 10^{8}$ & 285.7927
& 0.8 \\
EBLM J0500-35 & 8.28485 & 8.174±0.091 & $\le0.04518$ & 1.39±0.10 & 0.177±0.031 & 0.1273 & 7.78$\times 10^{4*}$ & 6.07$\times 10^{4*}$ & 1.00$\times 10^{11}$ & 8.92$\times 10^{4}$ & 0.7 \\
EBLM J0502-38 & 3.2563 & 3.26±0.10 & $\le0.01014$ & 1.26±0.1 & 0.307±0.054 & 0.2437 & 8.98$\times 10^5$ & 1.49$\times 10^{7}$ & 3.42$\times 10^{7}$ & 4.59$\times 10^{6}$ & 0.4 \\
EBLM J0520-06 & 2.13151 & 2.20±0.01 & $\le0.04733$ & 1.18±0.11 & 0.50±0.11 & 0.4237 & 2.21$\times 10^{4*}$ & 3.52$\times 10^{5*}$ & 2.74$\times 10^{6}$ & 7.44$\times 10^{5}$ & 0.5 \\
EBLM J0526+04 & 4.03101 & 4.59±0.01 & $\le0.00477$ & 1.17±0.1 & 0.193±0.03 & 0.165 & 1.05$\times 10^3$ & 1.19$\times 10^{3}$ & 6.53$\times 10^{8}$ & 5.26$\times 10^{3}$ & 0.8 \\
EBLM J0546-18 & 3.19191 & 3.27±0.23 & $\le0.01541$ & 1.21±0.09 & 0.231±0.04 & 0.1909 & 7.05$\times 10^5$ & 9.18$\times 10^{6*}$ & 8.34$\times 10^{7}$ & 2.26$\times 10^{6}$ & 0.5 \\
EBLM J0606-19 & 1.959995 & 1.96±0.04 & $\le0.0266$ & 1.2±0.1 & 0.29±0.052 & 0.2417 & 4.85$\times 10^{4*}$ & 4.15$\times 10^{5*}$ & 1.43$\times 10^{6*}$ & 8.88$\times 10^{5}$ & 0.5 \\
EBLM J0610-52 & 2.41699 & 2.43±0.18 & $\le0.00364$ & 1.47±0.11 & 0.60±0.12 & 0.4082 & 2.74$\times 10^{8*}$ & 5.32$\times 10^{9*}$ & 8.42$\times 10^{6}$ & 9.17$\times 10^{9}$ & 0.0 \\
EBLM J0621-46 & 1.55083 & 1.56±0.04 & $\le0.04875$ & 1.19±0.1 & 0.221±0.058 & 0.1857 & 3.46$\times 10^4$ & 1.81$\times 10^{5}$ & 7.74$\times 10^{4}$ & 6.78$\times 10^{5}$ & 0.4 \\
EBLM J0621-50 & 4.96384 & 5.11±0.08 & $\le0.00192$ & 1.23±0.1 & 0.42±0.06 & 0.3415 & 1.36$\times 10^5$ & 3.63$\times 10^{6*}$ & 1.40$\times 10^{9}$ & 8.95$\times 10^{5}$ & 0.7 \\
EBLM J0623-27 & 5.77793 & 5.82±0.05 & 0.05711±0.00121 & 1.18±0.08 & 0.308±0.038 & 0.261 & 9.84$\times 10^3$ & 1.11$\times 10^{1*}$ & 9.97$\times 10^{8}$ & 2.93$\times 10^{3}$ & 0.9 \\
EBLM J0625-43 & 3.96899 & 3.98±0.052 & $\le0.00748$ & 1.16±0.08 & 0.291±0.042 & 0.2509 & 5.55$\times 10^5$ & 1.11$\times 10^{7*}$ & 1.26$\times 10^{8}$ & 1.31$\times 10^{6}$ & 0.6 \\
EBLM J0640-27 & 2.92158 & 3.08±0.08 & $\le0.01066$ & 1.2±0.1 & 0.141±0.027 & 0.1175 & 1.61$\times 10^4$ & 5.73$\times 10^{4*}$ & 3.14$\times 10^{8}$ & 8.70$\times 10^{4}$ & 0.7 \\
EBLM J0649-27 & 4.30805 & 4.449±0.056 & $\le0.00233$ & 1.45±0.13 & 0.371±0.061 & 0.2559 & 1.42$\times 10^{3*}$ & 3.60$\times 10^{3*}$ & 2.72$\times 10^{9}$ & 7.61$\times 10^{3}$ & 0.4\\
EBLM J0659-61 & 4.23505 & 4.308±0.066 & $\le0.00139$ & 1.16±0.09 & 0.456±0.064 & 0.3931 & 1.50$\times 10^{1*}$ & 1.61 $\times 10^*$ & 5.07$\times 10^{8}$ & 7.86 $\times 10$ & 0.8 \\
EBLM J0916-35 & 2.58835 & 2.46±0.02 & $\le0.02226$ & 1.61±0.15 & 0.349±0.079 & 0.2168 & 7.74$\times 10^{9*}$ & 7.54$\times 10^{10*}$ & 6.50$\times 10^{6}$ & 1.03$\times 10^{11}$ & 0.0 \\
EBLM J0941-31 & 5.54563 & 5.38±0.04 & 0.20055±0.00172 & 1.19±0.1 & 0.218±0.031 & 0.1832 & 3.20$\times 10^{7*}$ & 9.33$\times 10^{8*}$ & 3.64$\times 10^{9*}$ & 8.90$\times 10^{6}$ & 0.1 \\
EBLM J0948-08 & 5.3798 & 5.36±0.01 & 0.04926±0.00019 & 1.41±0.12 & 0.675±0.096 & 0.4787 & 1.00$\times 10^7$ & 5.15$\times 10^{8}$ & 2.44$\times 10^{8}$ & 5.69$\times 10^{7}$ & 0.3 \\
EBLM J1013+01 & 2.89228 & 3.30±0.09 & $\le0.00885$ & 0.96±0.08 & 0.168±0.027 & 0.175 & 4.62$\times 10^5$ & 6.10$\times 10^{6}$ & 1.34$\times 10^{8}$ & 4.55$\times 10^{5}$ & 0.7 \\
EBLM J1016-42 & 4.3786 & 5.49±0.01 & $\le0.00462$ & 1.2±0.1 & 0.173±0.027 & 0.1442 & 1.11$\times 10^7$ & 2.21$\times 10^{8}$ & 1.90$\times 10^{9}$ & 8.68$\times 10^{6}$ & 0.4 \\
EBLM J1034-29 & 2.17426 & 2.18±0.02 & $\le0.007$ & 1.44±0.11 & 0.151±0.034 & 0.1049 & 1.60$\times 10^{7*}$ & 6.64$\times 10^{7*}$ & 4.18$\times 10^{7}$ & 1.05$\times 10^{8}$ & 0.2 \\
EBLM J1037-25 & 4.93656 & 4.94±0.07 & 0.12136±0.0005 & 1.26±0.1 & 0.26±0.038 & 0.2063 & 1.80$\times 10^{6*}$ & 3.52$\times 10^{7*}$ & 8.28$\times 10^{8}$ & 2.51$\times 10^{6}$ & 0.5 \\
EBLM J1037-45 & 1.59391 & 1.59±0.03 & $\le0.19367$ & 1.3±0.13 & 0.257±0.089 & 0.1977 & 1.70$\times 10^{4*}$ & 7.69$\times 10^{4*}$ & 7.16$\times 10^{5}$ & 7.08$\times 10^{5}$ & 0.4 \\
EBLM J1055-39 & 1.3516 & 1.37±0.05 & $\le0.57436$ & 1.33±0.12 & 0.249±0.115 & 0.1872 & 5.97$\times 10^4$ & 2.47$\times 10^{5}$ & 3.02$\times 10^{5}$ & 3.29$\times 10^{6}$ & 0.3  \\
EBLM J1116-01 & 4.742 & 4.74±0.13 & $\le0.00111$ & 1.17±0.08 & 0.59±0.075 & 0.5043 & 1.13$\times 10^{5*}$ & 4.51$\times 10^{6*}$ & 4.81$\times 10^{8}$ & 9.59$\times 10^{5}$ & 0.7 \\
EBLM J1141-37 & 5.14769 & 5.17±0.13 & $\le0.00126$ & 1.22±0.1 & 0.354±0.05 & 0.2902 & 1.10$\times 10^7$ & 4.47$\times 10^{8}$ & 3.81$\times 10^{8}$ & 1.26$\times 10^{7}$ & 0.3 \\
EBLM J1208-29 & 2.67598 & 2.68±0.04 & 0.17979±0.01614 & 1.6±0.11 & 0.25±0.058 & 0.1563 & 2.84 $\times 10^{10*}$ & 2.38$\times 10^{11*}$ & 2.58$\times 10^{7}$ & 1.87$\times 10^{11}$ & -0.1 \\
EBLM J1227-43 & 1.79554 & 1.80±0.03 & $\le0.06295$ & 1.3±0.11 & 0.267±0.070 & 0.2054 & 4.78$\times 10^{5*}$ & 3.66$\times 10^{6*}$ & 8.47$\times 10^{4*}$ & 5.78$\times 10^{6}$ & -1.3 \\
EBLM J1233-20 & 4.44384 & 4.70±0.07 & $\le0.01698$ & 1.14±0.08 & 0.198±0.028 & 0.1737 & 4.07$\times 10^{6*}$ & 8.19$\times 10^{7*}$ & 1.12$\times 10^{9*}$ & 2.99$\times 10^{6}$ &  0.6\\
EBLM J1243-50 & 1.54691 & 1.61±0.01 & $\le0.01709$ & 1.16±0.1 & 0.18±0.045 & 0.1552 & 2.22$\times 10^{4*}$ & 9.35$\times 10^{4*}$ & 2.18$\times 10^{6}$ & 3.59$\times 10^{5}$ & 0.6 \\
EBLM J1350-31 & 3.12433 & 3.16±0.03 & $\le0.01023$ & 1.17±0.08 & 0.378±0.061 & 0.3231 & 2.51$\times 10^5$ & 5.25$\times 10^{6}$ & 3.03$\times 10^{8}$ & 1.62$\times 10^{6}$ & 0.5  \\
EBLM J1418-29 & 8.69122 & 7.84±0.12 & 0.17867±0.00492 & 1.03±0.08 & 0.327±0.036 & 0.3175 & 2.99$\times 10^{5*}$ & 7.02$\times 10^{6*}$ & 1.02$\times 10^{10}$ & 1.34$\times 10^{5}$ & 1.0 \\
EBLM J1552-26 & 1.9492 & 1.99±0.02 & $\le0.02069$ & 1.2±0.12 & 0.155±0.039 & 0.1292 & 1.40$\times 10^{5*}$ & 6.65$\times 10^{5*}$ & 1.73$\times 10^{7*}$ & 9.12$\times 10^{5}$ & 0.5 \\
EBLM J1907-45 & 5.6402 & 5.66±0.02 & $\le0.01802$ & 1.02±0.09 & 0.311±0.039 & 0.3049 & 6.18$\times 10^{5*}$ & 2.02$\times 10^{7*}$ & 7.79$\times 10^{8}$ & 7.62$\times 10^{5}$ & 1.0 \\
EBLM J1910-35 & 4.08311 & 4.08±0.04 & $\le0.01682$ & 1.09±0.08 & 0.287±0.04 & 0.2633 & 1.00$\times 10^{6*}$ & 2.57$\times 10^{7*}$ & 1.51$\times 10^{8*}$ & 1.41$\times 10^{6}$ & 0.6 \\
EBLM J2001-36 & 2.75353 & 2.75±0.13 & $\le0.00394$ & 1.19±0.09 & 0.315±0.056 & 0.2647 & 1.18$\times 10^{5*}$ & 1.58$\times 10^{6*}$ & 9.70$\times 10^{6}$ & 9.71$\times 10^{5}$ & 0.6 \\
EBLM J2011-71 & 5.8727 & 5.90±0.18 & 0.03099±0.03112 & 1.41±0.13 & 0.285±0.042 & 0.2021 & 7.09$\times 10^{6*}$ & 1.50$\times 10^{8}$ & 1.93$\times 10^{9}$ & 2.23$\times 10^{7}$ & 0.4 \\
EBLM J2017-59 & 1.49548 & 1.51±0.07 & $\le0.15251$ & 1.2±0.12 & 0.138±0.058 & 0.115 & 8.83$\times 10^4$ & 2.88$\times 10^{5}$ & 4.89$\times 10^{6}$ & 7.95$\times 10^{5}$ & 0.3 \\
EBLM J2025-45 & 6.19199 & 6.43±0.08 & 0.12642±0.00049 & 0.96±0.06 & 0.218±0.022 & 0.2271 & 8.10$\times 10^6$ & 3.25$\times 10^{8}$ & 5.32$\times 10^{9}$ & 1.62$\times 10^{6}$ & 0.9 \\
EBLM J2107-39 & 3.9618 & 3.99±0.06 & $\le0.012$ & 1.2±0.08 & 0.253±0.038 & 0.2108 & 3.20$\times 10^6$ & 6.60$\times 10^{7}$ & 2.41$\times 10^{8}$ & 4.94$\times 10^{6}$  & 0.4\\
EBLM J2153-55 & 8.54481 & 10.15±0.11 & $\le0.00381$ & 1.04±0.07 & 0.316±0.03 & 0.3038 & 4.31$\times 10^{7*}$ & 3.98$\times 10^{9*}$ & 1.12$\times 10^{10}$ & 8.14$\times 10^{6}$ & 0.7 \\
 \hline
     \end{tabular}
     }
 \end{table*}

\begin{table*} 
    \centering
    \caption{}
    \resizebox{1.05\linewidth}{!}{
    \begin{tabular}{|l|l|l|l|l|l|l|l|l|l|l|l|}
    \hline
       \textbf{EBLM Name} & $\mathbf{P_{\rm orb}}$ & \textbf{Nominal} $\mathbf{P_{\rm rot}}$ & \textbf{Eccentricity} & $\mathbf{M_{\rm A}}$ & $\mathbf{M_{\rm B}}$ & \textbf{q} & $\tau_{\rm s}$ & $\tau_{\rm e, A}$ & $\tau_{\rm e, B}$ & $\mathbf{Q'}$ & $\log_{10}\left(\frac{\rm Age}{\rm Gyr}\right)$ \\ 
        & \textbf{[days]}&\textbf{[days]}& &\textbf{[$\mathbf{M_\odot}$]}&\textbf{[$\mathbf{M_\odot}$]}& & $\mathbf{[yr]}$ & $\mathbf{[yr]}$ & $\mathbf{[yr]}$ &  & \\ \hline
EBLM J2158-21 & 4.78197 & 4.818±0.028 & 0.03283±0.00567 & 1.1±0.08 & 0.27±0.037 & 0.2455 & 4.76$\times 10^{6*}$ & 7.39$\times 10^7*$ & 5.19$\times 10^{8*}$ & 2.04 $\times 10^6$ & 0.6 \\ 
EBLM J2210-48 & 2.8201 & 2.892±0.156 &
0±0.00187 & 1.37±0.11 & 0.363±0.069 & 0.265 & 4.59$\times 10^5$ & 2.07$\times 10^7*$ & 8.43$\times 10^6$  & 1.92 $\times 10^7$ & 0.3 \\ 
EBLM J2232-31 & 3.14153 & 3.14±0.045 & $\le0.01841$ & 1.25±0.1 & 0.215±0.039 & 0.172 & 1.55$\times 10^6$ & 1.82$\times 10^7$ & 1.04$\times 10^8$  & 4.50 $\times 10^6$ &  0.4 \\
EBLM J2236-36 & 3.06717 & 3.05±0.064 & $\le0.00353$ & 1.24±0.09 & 0.267±0.047 & 0.2153 & 2.22$\times 10^{5*}$ & 3.13$\times 10^6*$ & 3.69$\times 10^7$  & 1.41 $\times 10^6$ & 0.5 \\ 
EBLM J2308-46 & 2.19922 & 2.21±0.065 & $\le0.01384$ & 1.23±0.1 & 0.181±0.04 & 0.1472 & 3.40$\times 10^{4*}$ & 1.63 $\times 10^5*$ & 2.01$\times 10^7$  & 2.74 $\times 10^5$ & 0.6 \\ 
EBLM J2309-67 & 1.95519 & 2.005±0.005 & $\le0.01384$ & 1.23±0.1 & 0.181±0.04 & 0.1472 & 1.38$\times 10^{5*}$ & 7.49$\times 10^5*$ & 9.30$\times 10^{6*}$  & 1.28 $\times 10^6$ &  0.4 \\ 
EBLM J2349-32 & 3.54967 & 4.232±0.048 & $\le0.00235$ & 1.19±0.08 & 0.195±0.031 & 0.1639 & 1.85$\times 10^6$ & 2.87$\times 10^7$ & 3.09$\times 10^8$  & 3.45 $\times 10^6$ & 0.5 \\ \hline
\end{tabular}
}
\end{table*}


\bsp	
\label{lastpage}
\end{document}